\documentclass{article}

\usepackage[preprint]{neurips_2026}

\usepackage[utf8]{inputenc} 
\usepackage[T1]{fontenc}    
\usepackage{hyperref}       
\usepackage{url}            
\usepackage{booktabs}       
\usepackage{amsfonts}       
\usepackage{nicefrac}       
\usepackage{microtype}      
\usepackage{xcolor}         

\usepackage{blindtext} 
\usepackage{subcaption} 
\usepackage{wrapfig, lipsum,booktabs}
\usepackage{algorithm,algpseudocode}
\usepackage{enumitem}
  \setlist{leftmargin=*}
\usepackage{graphicx}
  \graphicspath{ {./figs/} }
\usepackage{lineno}
\usepackage{tikz}
\usetikzlibrary{positioning, arrows.meta, calc, decorations.pathreplacing, fit, shapes.geometric, positioning, backgrounds}
\usepackage[T1]{fontenc}
\usetikzlibrary{positioning}
\usepackage{YK}

\newcount\Comments  
\newcommand{\kibitz}[2]{\ifnum\Comments=1\textcolor{#1}{#2}\fi}
\usepackage{xcolor}
\definecolor{agentblue}{HTML}{4A90D9}
\definecolor{promptgreen}{HTML}{5CB85C}
\definecolor{llmpurple}{HTML}{8E44AD}
\definecolor{validgold}{HTML}{F0AD4E}
\definecolor{successgreen}{HTML}{27AE60}
\definecolor{failred}{HTML}{E74C3C}
\definecolor{infogray}{HTML}{ECF0F1}
\definecolor{contextblue}{HTML}{3498DB}
\usepackage[T1]{fontenc}
\usepackage{lmodern}
\definecolor{darkgreen}{rgb}{0,0.4,0}
\definecolor{purple}{rgb}{1,0,1}

\usepackage{listings}
\usepackage{natbib}
\usepackage{listings}
\usepackage{multirow}
\usepackage{microtype}      
\usepackage{colortbl}
\usepackage{bigstrut}
 \usepackage{xcolor}         
\usepackage{graphicx}
\usepackage{rotating} 
\usepackage{pdflscape} 
\usepackage{adjustbox}
\usepackage{comment}
\usepackage[font=small]{caption}
\usepackage{tcolorbox}
\tcbuselibrary{skins, breakable}
\newtcolorbox{mybox}{colback=blue!5!white,colframe=blue!75!black}
\definecolor{Gray}{gray}{0.95}
\definecolor{DarkGray}{gray}{0.5}
\definecolor{LightCyan}{rgb}{0.88,1,1}
\definecolor{bisque}{rgb}{1.0, 0.89, 0.77}
\definecolor{blanchedalmond}{rgb}{1.0, 0.92, 0.8}
\definecolor{cosmiclatte}{rgb}{1.0, 0.97, 0.91}
\definecolor{cornsilk}{rgb}{1.0, 0.97, 0.86}
\title{The Open Source Economic Index of AI Adoption and Capability}

\author{%
  Seamus Somerstep\thanks{Work partially completed while an intern at the AT\&T Chief Data Office} \\
  University of Michigan \\
  \texttt{smrstep@umich.edu} \\
  \And
  Aritra Guha \\
  AT\&T Chief Data Office\\
  \AND
  Divesh Srivastava \\
  AT\&T Chief Data Office\\
  \And
  Yuekai Sun \\
  University of Michigan \\
  IFM-MBZUAI \\
}

\begin{document}

\maketitle

\begin{abstract}
   We work towards measuring both AI adoption and the capability of AI to perform discrete labor tasks across various occupations\footnote{Code available at \url{https://github.com/UMich-FATML/open-econ-index}}. To  measure adoption, we develop an open-source economic index that uses publicly available user-LLM chat data and O*NET tasks to replicate studies produced by frontier AI labs, finding that occupations in the finance, computer science, and arts sectors are those with the highest adoption rates. To measure capabilities, we build a system that generates benchmark scenarios grounded in O*NET occupations, tasks, and model-context-protocol (MCP) servers. We test Kimi-k2.5 with an OpenAI agents SDK harness on scenarios across 9 occupations that appear frequently in our index, finding that AI correctly executes high-level workflows but often errs in the granular details (such as specific tool calls used).
\end{abstract}

\lstdefinestyle{promptstyle}{
  basicstyle=\ttfamily\footnotesize,
  breaklines=true,
  breakatwhitespace=false,
  columns=fullflexible,
  keepspaces=true,
  frame=single,
  framesep=6pt,
  backgroundcolor=\color{gray!8},
  rulecolor=\color{gray!50},
  xleftmargin=6pt,
  xrightmargin=6pt,
  aboveskip=8pt,
  belowskip=8pt,
}
\lstdefinestyle{conversationstyle}{
  basicstyle=\ttfamily\footnotesize,
  breaklines=true,
  breakatwhitespace=false,
  columns=fullflexible,
  keepspaces=true,
  frame=single,
  framesep=6pt,
  backgroundcolor=\color{blue!4},
  rulecolor=\color{blue!30},
  xleftmargin=6pt,
  xrightmargin=6pt,
  aboveskip=8pt,
  belowskip=8pt,
}

\section{Introduction}
Global private investment in generative AI grew to \$33.9 billion in 2024, while 78\% of organizations reported AI usage \citep{stanford2025aiindex}. Despite this, empirical evidence regarding its granular impact on specific work tasks remains opaque. In combination with the deep embedding of AI in the economy, this has the potential to create "excessive automation" or the automation of work tasks that displace labor and depress wages without generating significant gains in productivity \citep{acemoglu2021harms}. 

To alleviate this risk, efforts to snapshot the capabilities of AI in the workforce have increased. Human surveys \citep{humlum2024adoption, bick2024rapid} can provide an estimate but suffer from under-reporting bias \citep{ling2025underreporting}. Controlled experiments \citep{peng2023impact, noy2023experimental} can alleviate this issue but offer insights that are limited to singular occupations. For a broader picture, \citet{handa2025economictasksperformedai, NBERw34255} and \citet{tomlinson2025working} have developed systems to map user-LLM chats to O*NET \citep{onet2025online} tasks or work activities; creating aggregate statistics of what labor activities individuals use AI for. While this method has illuminated patterns in AI usage, it suffers from two shortcomings. First, prior work has used proprietary chat data, making it difficult to corroborate findings. Second, simply categorizing chats into an occupational task or work activity does not measure the \emph{performance} of the LLM in that task or activity. \citet{NBERw34255} create a preliminary analysis by analyzing user chats for sentiment; however, like other works, the chats are proprietary, and sentiment may not be uniquely determined by the model's accuracy. 

\begin{figure}[t]
    \centering
    \includegraphics[width=0.7\linewidth]{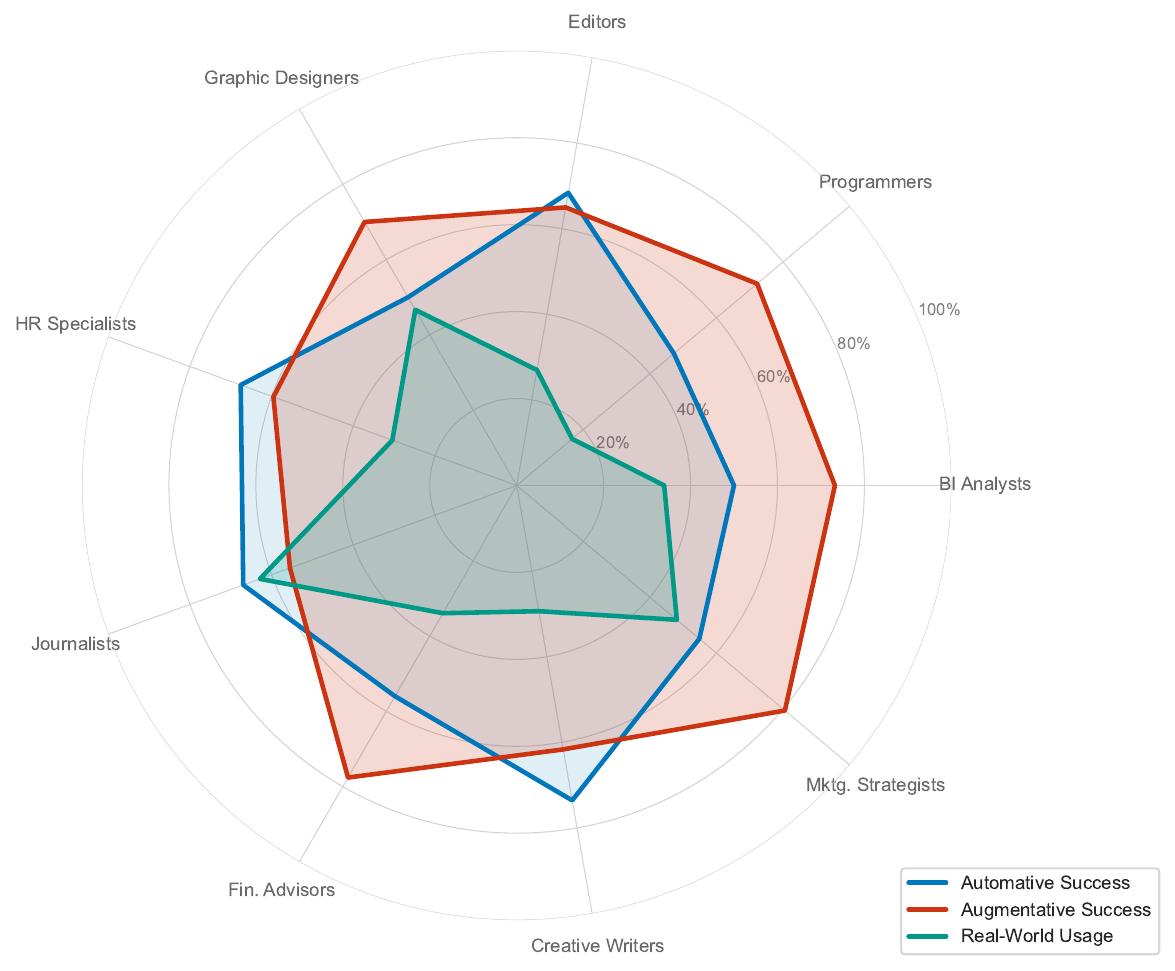}
    \caption{Comparing AI agent capability and real-world AI adoption across nine occupations.}
    \label{fig:hookplot}
\end{figure}

Other works utilize human generated and graded deliverables to measure LLM economic value. \citet{patwardhan2025gdpval} design GDPVal, a benchmark constructed from 1,320 tasks across 44 occupations representing the top 9 sectors of U.S. GDP. They find that AI has a win-rate exceeding 50\% against experts in certain occupations. \citet{mazeika2025rli} produces contrasting results; they crowd-source complex projects from free-lancers and find that current agents can only automate less than 3\% of projects. The staunch disagreement is partially due to the creation of custom tasks by \citet{patwardhan2025gdpval} and \citet{mazeika2025rli}, and it reveals a need for further investigation.

In our view, obtaining a snapshot of the labor impacts of AI involves seeking answers to two questions.
\begin{enumerate}
    \item Which occupations are using AI?
    \item Do these occupations benefit from the use of GenAI, and are they at risk of significant job loss? 
\end{enumerate}

To help provide an answer to these questions, we build two systems for studying AI adoption and capabilities. First, we construct
an open source economic index that measures the occupational sectors with high LLMs utilization (Figure \ref{fig: representation}), and the depth of AI penetration into the economy (Figure \ref{fig: depth}). Throughout, we will discuss how these findings align  with prior work \citep{NBERw34255, handa2025economictasksperformedai, tomlinson2025working}. Second, we build an economic benchmark that confers two advantages over GDPval \citep{patwardhan2025gdpval} and the remote labor index 
\citep{mazeika2025rli}. We test the agents on O*NET tasks, removing the ambiguity that arises from creating custom tasks. Furthermore, our system is neither human generated nor evaluated, allowing us to create a greater number of benchmark scenarios per occupation. Using our benchmark, we evaluate the performance of agents in scenarios covering 9 occupations. 
Figure \ref{fig:hookplot} encapsulates our findings; we plot the AI agent capability and real-world AI adoption across nine occupations. The automative plot (blue) shows the share of single-turn scenarios achieving perfect workflow completion, reflecting autonomous task execution.  The Augmentative plot (red) shows the share of multi-turn scenarios achieving perfect completion, reflecting collaborative task execution with a simulated user. Real-world task usage (teal) shows the share of scenarios whose constituent O*NET tasks appear at least four times in our open economic index. Overall, we find that theoretical capabilities outpace adoption in the index, but complete occupational replacement risks may be minimal as user-AI collaboration outpaces purely autonomous task execution. Further evidence of the need for human oversight is given in Figure \ref{fig:hruv2}, where we see the tested agent still has non-negligible tool calling error and hallucination rates.

\subsection{Related Work}
Our work builds on task-centric labor analysis \citep{autor2003skill, autor2013task, acemoglu2018race}, empirical studies of AI adoption \citep{humlum2024adoption, bick2024rapid, noy2023experimental, peng2023impact}, and agentic benchmarks such as BFCL \citep{patil2025bfcl}, MCPBench \citep{wang2025mcpbench}, and $\tau$-bench \citep{yao2025taubench}. A full review of related work is provided in Appendix \ref{app:related-work}.

\section{Open Source Measurement of AI Adoption}

To create an open source index of how individuals utilize large language models across the economy, we design a semantic search system that utilizes gpt-oss-120b \citep{openai2025gptoss120bgptoss20bmodel} and the Qwen3-Embedder \citep{zhang2025qwen3embeddingadvancingtext} to map user-LLM conversations to tasks in the O*NET database. 

\begin{table}[h!]
    \centering
    \label{tab:1x3_bold_headers}
    \begin{tabular}{|c|c|}
        \hline
        \textbf{Occupationally Relevant} & \textbf{Occupationally Irrelevant} \\ 
        \hline
        789768 (47.2 \%) & 884548 (52.7 \%)\\
        \hline
    \end{tabular}
    \caption{WildChat chats filtered for occupational relevancy}
    \label{tab: OR}
\end{table}

\subsection{Data and Methodology}
 We begin with the  WildChat-4.8M \citep{zhao2024wildchat} dataset available on HuggingFace. WildChat-4.8M is a collection of 3,199,860 non-toxic inputs and responses between human users and various versions of ChatGPT. Exchanges may be over 25 turns, and model types include gpt-3.5-turbo, gpt-4, and even reasoning models such as o3-mini. The flip side of the index is the O*NET database; a taxonomy that organizes occupations according to their tasks, required knowledge levels, and skills. Each occupation is associated with tasks that are further embedded in a four level hierarchy that also includes 2,087 detailed work activities (DWAs), 332 intermediate work activities (IWAs), and 41
work activities (WAs).

To make the raw chat data more usable for measuring work-related activity, we apply three data filters. The first and second are simple language and uniqueness filters that isolate the unique English chats, leaving us with 1674316 total chats for analysis. 
Similar to \citet{handa2025economictasksperformedai} and \citet{NBERw34255}, many of the English chats in the WildChat dataset are likely not related to any type of work activity. 
It is unlikely that these can be given any meaningful mapping to an O*NET task. To prevent this from decreasing the accuracy of our index, we design a simple filter for ``Occupational Relevancy". To perform the filter, each chat is fed to gpt-oss-120b along with a prompt that asks if the user's request is "occupationally relevant (related to work, professional tasks, or career activities)". The results, given in Table \ref{tab: OR}, are remarkably similar to the June 2024 results presented by \citet{NBERw34255}. WildChat chats were collected by providing free ChatGPT access to users and were subsequently filtered for toxicity. It is worth noting that \citet{NBERw34255} also reported increased relevancy rates in 2025, suggesting that LLM usage practices are not static.

\begin{figure}[h]
\footnotesize
\centering
\begin{tikzpicture}[
    node distance=0.45cm,
    box/.style={draw, rounded corners, text width=7.5cm, align=left, font=\footnotesize, inner sep=6pt},
    arrow/.style={-{Stealth[length=2.5mm]}, thick},
    label/.style={font=\scriptsize\bfseries, text=gray}
]
\node[box, fill=blue!5] (chat) {
    \textbf{User:} Write a VBA to calculate RSI\ldots \\[2pt]
    \textbf{Asst:} To calculate RSI using VBA\ldots
};
\node[label, left=0.2cm of chat.west, anchor=east] {Chat};
\node[box, fill=green!5, below=of chat] (summary) {
    Write a VBA script that calculates RSI.
};
\node[label, left=0.2cm of summary.west, anchor=east] {Summary};
\draw[arrow] (chat) -- node[right, font=\scriptsize, text=gray] {LLM} (summary);
\node[box, fill=orange!5, below=of summary] (tasks) {
    (1)~Analyze data using statistical software.\\[1pt]
    (2)~Clean and manipulate raw data.
};
\node[label, left=0.2cm of tasks.west, anchor=east] {Tasks};
\draw[arrow] (summary) -- node[right, font=\scriptsize, text=gray] {Embed + Filter} (tasks);
\node[box, fill=red!5, below=of tasks] (occ) {
    \textbf{Data Scientists} (SOC 15-2051.00)
};
\node[label, left=0.2cm of occ.west, anchor=east] {Occupation};
\draw[arrow] (tasks) -- node[right, font=\scriptsize, text=gray] {O*NET Lookup} (occ);
\end{tikzpicture}
\caption{Chat-to-occupation mapping pipeline.}
\label{fig:pipeline-example}
\end{figure}
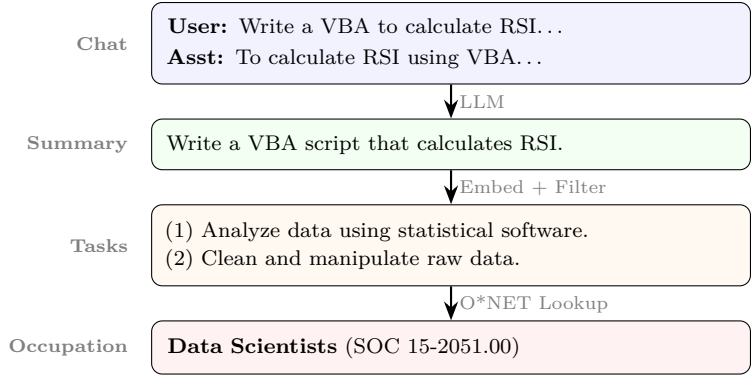

The second step in our analysis is to map each of the 789768 occupationally relevant chats to O*NET task(s). To do this, we implement the following three step semantic search process:
\begin{enumerate}
    \item For each occupationally relevant chat, we have gpt-oss-120b summarize the user's chat(s) into a single overarching request for the LLM.
    \item Using the Qwen3-Embedding-0.6B model, we create instruction-augmented embeddings of each chat summary, embeddings of each O*NET task, and map 3 tasks to each chat using a nearest neighbors search.
    \item Gpt-4o-mini serves as a filter for each (chat summary, 3 tasks) pair and is asked to keep any task where “the core skill or activity overlaps".
\end{enumerate}

As a model validation step for the chosen embeding model, we compute the cophenetic correlation between the embedding distance and the O*NET hierarchy graph distance. Graph distance is computed as the shortest path (hop count) in the undirected O*NET hierarchy.  In Figure \ref{fig:CPHN-H} we demonstrate the distance between two tasks with the dashed red path; cophenetic correlation is given over all $\binom{|\text{GWA+IWA+DWA+Task}|}{2}$ node pairs across all levels. Table \ref{tab:cophenetic-correlations-model} shows good results for Qwen embedding models of two sizes; both handily beat the model\footnote{Anthropic used \url{https://huggingface.co/sentence-transformers/all-mpnet-base-v2}.} \citet{handa2025economictasksperformedai} selected for the task.

\begin{figure}[h]
\centering
\begin{minipage}[c]{0.55\textwidth}
\centering
\resizebox{\textwidth}{!}{%
\begin{tikzpicture}[
    >=Stealth,
    every node/.style={font=\small},
    root/.style={circle, draw, fill=gray!30, minimum size=8mm, font=\footnotesize\bfseries},
    gwa/.style={circle, draw, fill=blue!20, minimum size=8mm, font=\footnotesize},
    iwa/.style={circle, draw, fill=green!20, minimum size=7mm, font=\footnotesize},
    dwa/.style={circle, draw, fill=orange!20, minimum size=7mm, font=\footnotesize},
    task/.style={circle, draw, fill=red!20, minimum size=6mm, font=\scriptsize},
    edge/.style={draw, thick, gray!60},
    highlight/.style={draw, very thick, red!70!black, dashed},
    level label/.style={font=\small\itshape, text=gray},
  ]
  \node[root] (root) at (0, 6) {$r$};
  \node[gwa] (g1) at (-3, 4.2) {$g_1$};
  \node[gwa] (g2) at ( 3, 4.2) {$g_2$};
  \node[iwa] (i1) at (-4.5, 2.4) {$i_1$};
  \node[iwa] (i2) at (-1.5, 2.4) {$i_2$};
  \node[iwa] (i3) at ( 1.5, 2.4) {$i_3$};
  \node[iwa] (i4) at ( 4.5, 2.4) {$i_4$};
  \node[dwa] (d1) at (-5.5, 0.8) {$d_1$};
  \node[dwa] (d2) at (-3.5, 0.8) {$d_2$};
  \node[dwa] (d3) at (-1.5, 0.8) {$d_3$};
  \node[dwa] (d4) at ( 1.5, 0.8) {$d_4$};
  \node[dwa] (d5) at ( 3.5, 0.8) {$d_5$};
  \node[dwa] (d6) at ( 5.5, 0.8) {$d_6$};
  \node[task] (t1) at (-6.2, -0.8) {$t_1$};
  \node[task] (t2) at (-4.8, -0.8) {$t_2$};
  \node[task] (t3) at (-3.5, -0.8) {$t_3$};
  \node[task] (t4) at (-1.5, -0.8) {$t_4$};
  \node[task] (t5) at ( 1.0, -0.8) {$t_5$};
  \node[task] (t6) at ( 2.2, -0.8) {$t_6$};
  \node[task] (t7) at ( 3.5, -0.8) {$t_7$};
  \node[task] (t8) at ( 5.5, -0.8) {$t_8$};
  \draw[edge] (root) -- (g1); \draw[edge] (root) -- (g2);
  \draw[edge] (g1) -- (i1); \draw[edge] (g1) -- (i2);
  \draw[edge] (g2) -- (i3); \draw[edge] (g2) -- (i4);
  \draw[edge] (i1) -- (d1); \draw[edge] (i1) -- (d2);
  \draw[edge] (i2) -- (d3); \draw[edge] (i3) -- (d4);
  \draw[edge] (i4) -- (d5); \draw[edge] (i4) -- (d6);
  \draw[edge] (d1) -- (t1); \draw[edge] (d1) -- (t2);
  \draw[edge] (d2) -- (t3); \draw[edge] (d3) -- (t4);
  \draw[edge] (d4) -- (t5); \draw[edge] (d4) -- (t6);
  \draw[edge] (d5) -- (t7); \draw[edge] (d6) -- (t8);
  \draw[highlight] (t2) -- (d1); \draw[highlight] (d1) -- (i1);
  \draw[highlight] (i1) -- (g1); \draw[highlight] (g1) -- (root);
  \draw[highlight] (root) -- (g2); \draw[highlight] (g2) -- (i3);
  \draw[highlight] (i3) -- (d4); \draw[highlight] (d4) -- (t5);
  \draw[decorate, decoration={brace, amplitude=5pt, mirror}, thick, red!70!black]
    ([yshift=-3pt]t2.south) -- ([yshift=-3pt]t5.south)
    node[midway, below=8pt, font=\small, text=red!70!black] {$d_{\mathrm{graph}}(t_2, t_5) = 8$ hops};
  \node[level label] at (7.2, 6) {Root};
  \node[level label] at (7.2, 4.2) {GWA};
  \node[level label] at (7.2, 2.4) {IWA};
  \node[level label] at (7.2, 0.8) {DWA};
  \node[level label] at (7.2, -0.8) {Task};
\end{tikzpicture}
}%
\end{minipage}
\hfill
\begin{minipage}[c]{0.42\textwidth}
\centering
\captionof{table}{Embedding cophenetic corr.}
\label{tab:cophenetic-correlations-model}
\vspace{4pt}
\small
    \begin{tabular}{|c|c|}
        \hline
        \textbf{Model} & \textbf{Pearson $r$} \\
        \hline
        all-mpnet-base-v2 & 0.2 \\
        \hline
        Qwen3-Emb.-0.6B & 0.25 \\
        \hline
        Qwen3-Emb.-8B & 0.31 \\
        \hline
    \end{tabular}
\end{minipage}
\caption{O*NET hierarchy with graph distance (left) and cophenetic correlations by embedding model (right).}
\label{fig:CPHN-H}
\end{figure}

This two step process is novel in the context of LLM economic index production. \citet{handa2025economictasksperformedai} create a custom taxonomy using clustering and perform an LLM powered tree search to map conversations to O*NET tasks. \citet{NBERw34255} and \citet{tomlinson2025working} only map chats to the IWA level of the O*NET hierarchy, losing the granularity of task-level mapping. 

\subsection{Results}

\begin{figure*}[t!]
\centering
\begin{subfigure}[t]{0.48\textwidth}
  \centering
  \includegraphics[width=\textwidth,height=7.5cm]{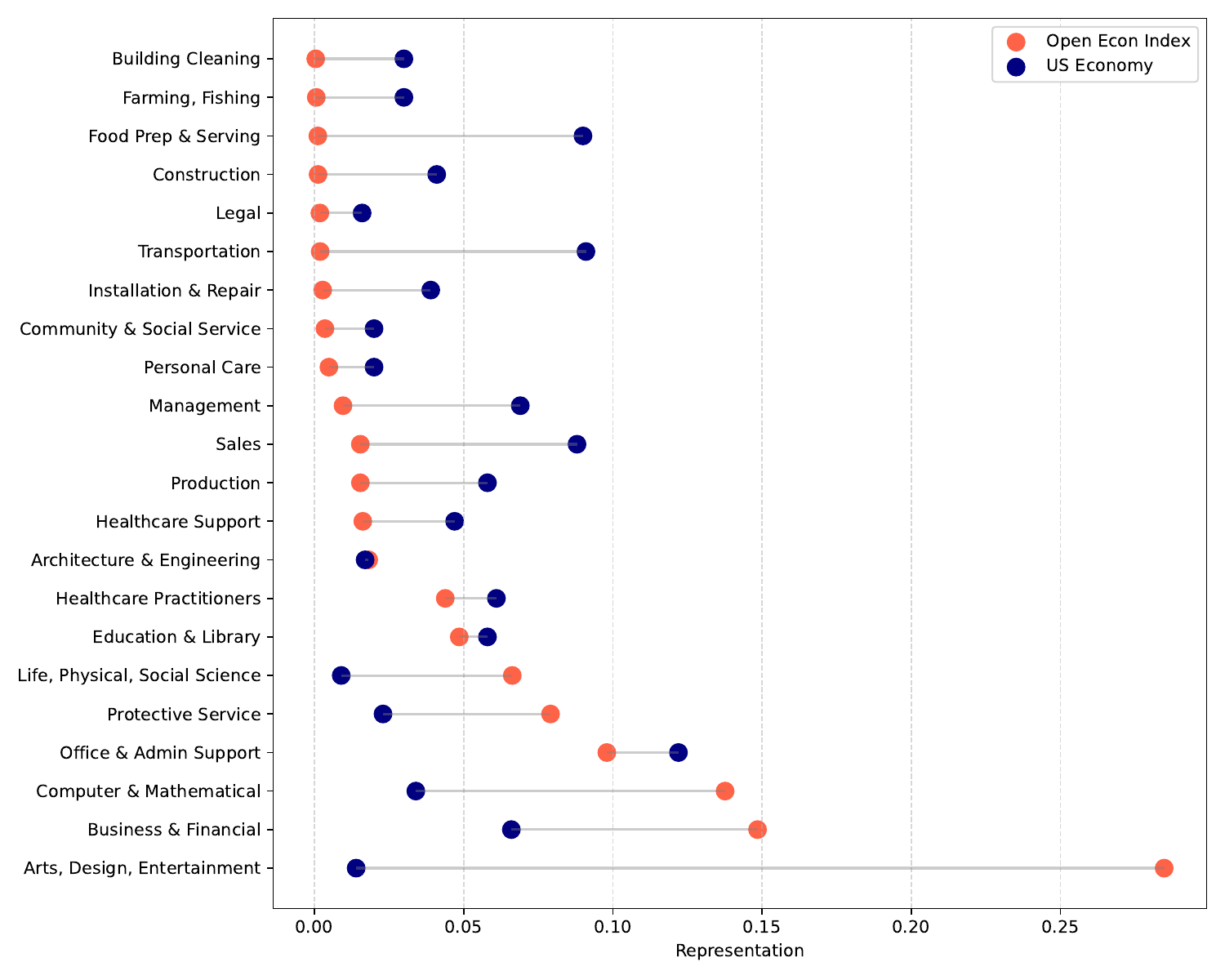}
  \caption{Comparison of occupational representation in WildChat data and the U.S. economy.}
  \label{fig: representation}
\end{subfigure}
\hfill
\begin{subfigure}[t]{0.48\textwidth}
  \centering
  \includegraphics[width=\textwidth,height=7.5cm]{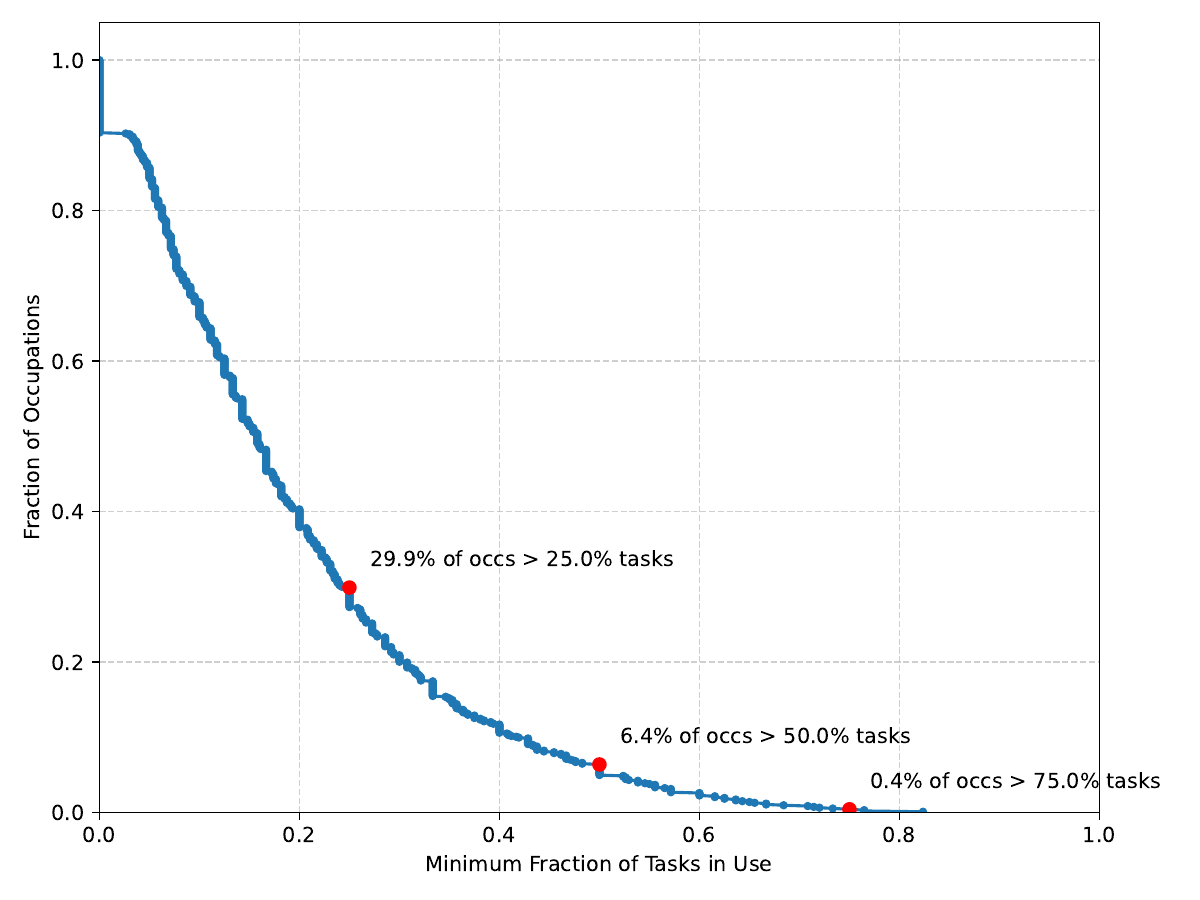}
  \caption{Depth of AI usage across occupations.}
  \label{fig: depth}
\end{subfigure}
\caption{Open Economic Index: Representation and Depth}
\end{figure*}
Of the 789768 occupationally relevant chats, 668380 chats have at least one represented task; with a split of 416,611 having 3 tasks, 169,835 having 2 tasks, and 81,934 having 1 task.
Upon mapping chats to O*NET tasks, we aggregate to the occupation level. For each chat, a given occupation is assigned a weight proportional to the count of its tasks within the range of that chat. The representation of a given occupation in the economic index is then calculated via a weighted sum. Overall, Figure \ref{fig: representation} (produced by isolating chats geo-located to the U.S) provides several insights\footnote{U.S. representation is from \citet{usbls}}.
Relative to the US economy, LLM usage is concentrated in the arts, business, financial work, programming, mathematics, and subject-specific sciences. This aligns with the LLM usage that \citet{handa2025economictasksperformedai} measured across the economy. 
Surprisingly, the white collar field with the lowest LLM adaptation is legal services, a conclusion that Anthropic also reached \citep{handa2025economictasksperformedai}. Compared to the Anthropic economic index, we find relatively less usage in educational tasks and more in protective service tasks, which is contrary to the findings in \citep{handa2025economictasksperformedai}. Protective services include translation, which accounts for the mappings in our chat database. Among the ``LLM prevalent" occupations, our index places the greatest weight on the Arts, while \citet{handa2025economictasksperformedai} places the greatest weight on computer science tasks.   

We can also use the index to assess the depth of AI penetration into the economy. \citet{handa2025economictasksperformedai} measure this by calculating the percentage of occupations that have at least $p$ tasks in the image of at least $t$ conversations, where $p$ can range from $0$ to $1$. We present the same calculation in Figure \ref{fig: depth} using our index, where we adjust the threshold to account for the smaller number of conversations present in our dataset. Compared to prior work, our results are bearish on AI depth, with only 0.4\% of occupations showing AI usage for at least 75\% of their tasks (as compared to the estimated 4\% of occupations given by \citet{handa2025economictasksperformedai}). 

\begin{figure}[t!]
\centering
\begin{subfigure}[t]{0.48\textwidth}
  \centering
  \includegraphics[width=\textwidth,height=8cm]{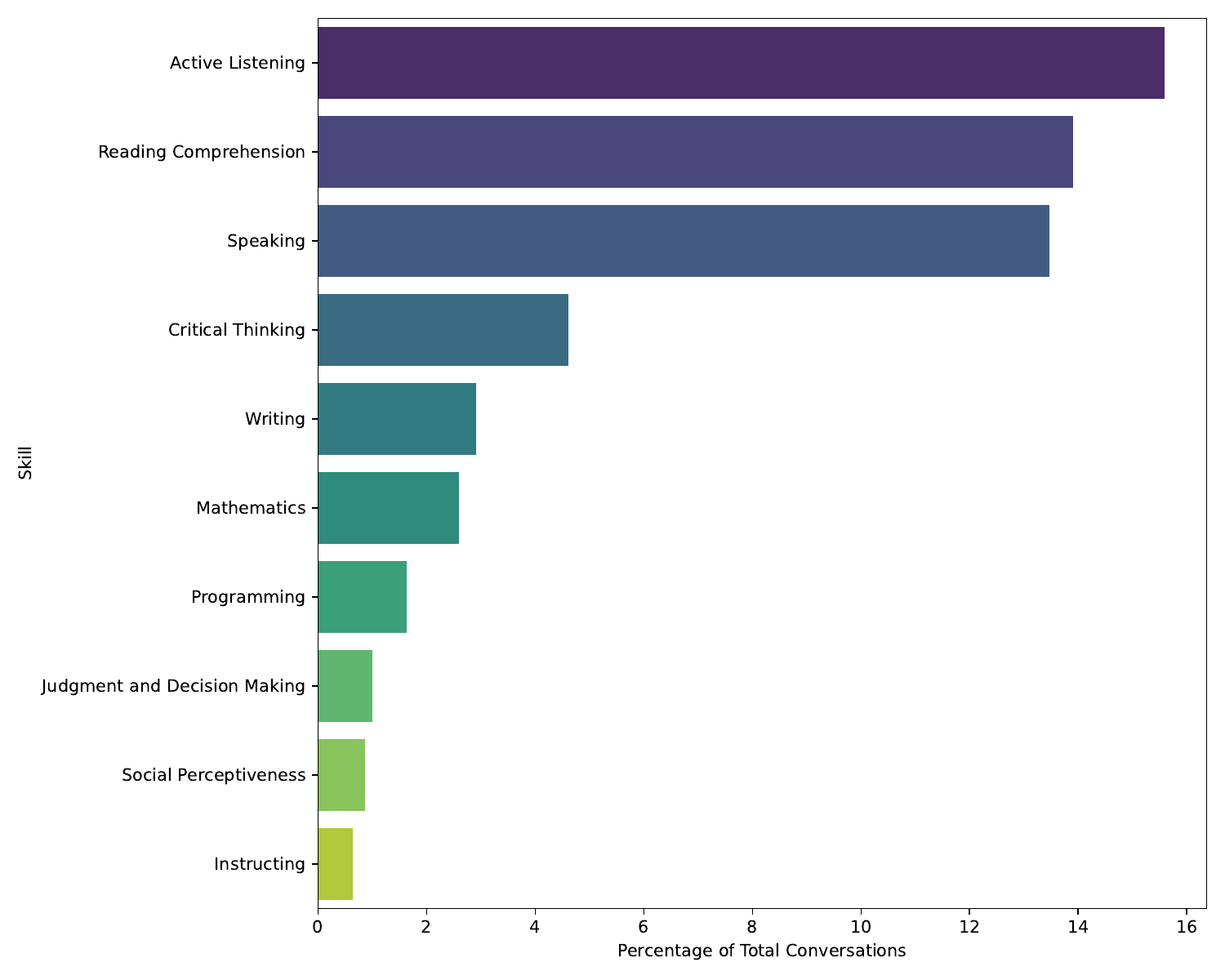}
  \caption{Distribution of important skills for occupations appearing in the WildChat dataset.}
  \label{fig:skill}
\end{subfigure}
\hfill
\begin{subfigure}[t]{0.48\textwidth}
  \centering
  \includegraphics[width=\textwidth,height=8cm]{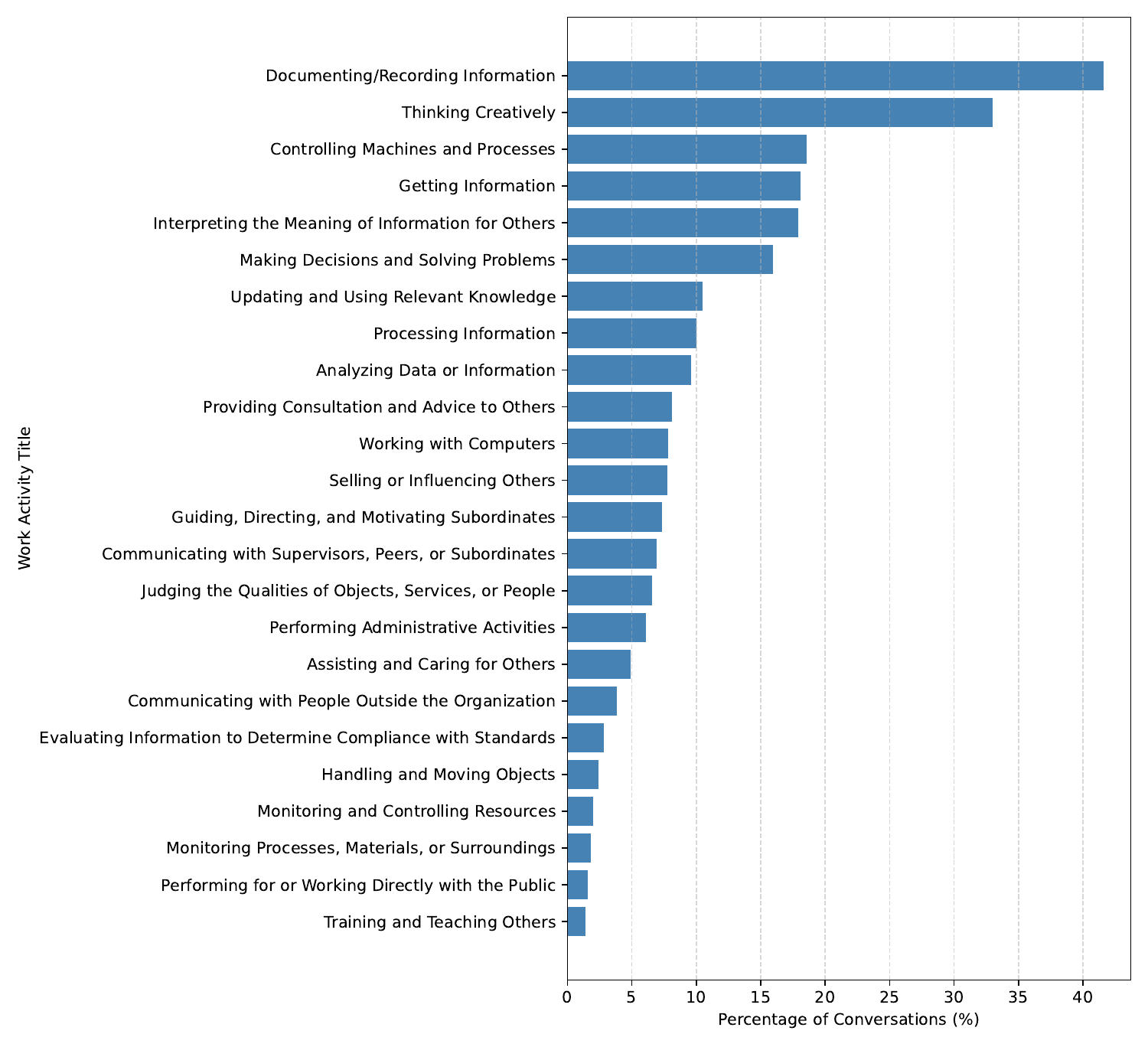}
  \caption{Work Activity Coverage across approximately 789,000 conversations.}
  \label{fig:WA}
\end{subfigure}
\caption{Skills and Work Activities in the Open Economic Index}
\end{figure}

The skills utilized in the tasks attempted by AI are also of interest. We infer skill usage using the O*NET data structure and our open economic index. O*NET rates every occupation on 35 specific skills (like "Critical Thinking," "Programming," or "Social Perceptiveness").
For a given skill, we calculate how many chats map to an occupation that rates that skill as having an importance of at least 4. Figure \ref{fig:skill} presents the skills most frequently required to perform the occupations listed in our open economic index. Cognitive skills such as writing, programming, reading comprehension, critical thinking, and active listening are dominant in the index. 

As with the Anthropic index \citep{handa2025economictasksperformedai}, our open source economic index can be compared with the index produced by OpenAI \citep{NBERw34255}. Their top appearing work activities include ``Thinking Creatively", ``Documenting/Recording Information", ``Making Decisions and Solving Problems", and ``Working with Computers". Figure \ref{fig:WA} provides the calculation we perform using WildChat data and our search system;  tasks matched to conversations are aggregated to Work Activities (WAs) using the O*NET Work Activities taxonomy. 
Except for working with computers, all top work activities from OpenAI appear within the top ten work activities we produce \citep{NBERw34255}. 
\section{Benchmarking the Economic Capabilities of Agents}
\label{sec: benchmark}
{ The open economic index characterizes which occupational tasks appear in real-world user–LLM interactions and how frequently they arise across occupations. However, the presence of a task in observed usage does not, by itself, indicate whether contemporary AI systems can reliably or autonomously perform that task. User interactions may reflect exploratory use, partial assistance, or heavily supervised workflows, and do not directly measure task completion quality or execution correctness.

To move beyond descriptive adoption and toward a functional assessment of AI capabilities, we require a complementary evaluation framework that can measure how agents perform on the same classes of tasks as observed in real-world usage. In particular, such a framework should (i) be grounded in the same occupational task taxonomy (O*NET) used by the index, (ii) evaluate multi-step task execution rather than isolated responses, and (iii) distinguish between autonomous execution and collaborative, user-steered workflows.
Motivated by these gaps, we next introduce an economic capability benchmark that translates O*NET tasks into tool-mediated workplace scenarios and directly evaluates agent performance on workflow completion, tool use, and grounding. We build a system to generate agentic evaluations that are grounded in O*NET data and real or virtual MCP servers.} We test agents in two modes: \emph{single-turn} (autonomous execution) and \emph{multi-turn} (with a simulated user providing collaborative steering).
\subsection{Scenario construction}
The benchmark scenarios are grounded in the O*NET database. The tool ecosystem for the benchmark consists of Smithery\footnote{\url{https://smithery.ai/}} MCP (Model Context Protocol) servers; we crawl every such server, producing an LLM generated analysis and quality assessment. Additionally, for stability and increased tool coverage, we implement a  benchmark mode that uses simulated MCP servers. Details are provided in Appendix \ref{app: servers}.

Beyond filtering, the second core challenge is connecting each O*NET task to the MCP servers whose tools are most likely to help accomplish that task. We perform this match semantically, using the server analysis generated in step one, via a Qwen3-Embedding model (thresholding a valid combination at a cosine similarity of 0.5). Scenarios are generated by enumerating (occupation, N (task, server) ) tuples and prompting an LLM to produce a grounded workplace request with a target workflow and a granular answer key of required tool calls, including both tool names and specified arguments. An example is given in Appendix \ref{app:scenario-examples}.

For multi-turn evaluation, we generate information-withheld variants where critical parameters are omitted from the request, requiring the agent to ask clarifying questions Appendix \ref{app:scenario-examples}. A simulated user agent provides answers and collaborative steering via role inversion (Figure \ref{fig:role inversion}, Appendix \ref{app: user}).
\subsection{Scenario evaluation}
Each trajectory is evaluated using LLM-as-judge across five dimensions (full methodology and rubrics in Appendix \ref{app:eval-methodology}).
\textbf{Tool Call Accuracy} (binary) measures whether the agent called correct tools with correct arguments, following the BFCL approach \citep{patil2025bfcl} with LLM-judged functional equivalence.
\textbf{Workflow Completion} (1--5, Table \ref{tab:workflow-rubric}) assesses whether the agent followed the intended multi-step tool sequence with proper inter-tool data flow.
\textbf{Grounding} (1--5, Table \ref{tab:grounding-rubric}) evaluates whether the agent's claims are supported by tool outputs rather than hallucinated ones.

For multi-turn trajectories, we add two dimensions alongside tool calling and grounding: \textbf{Autonomy} (1--5, Table \ref{tab:rubric-autonomy}) measures how independently the agent completes the workflow without user steering (analog to workflow completion), and \textbf{Follow-up Quality} (1--5, Table \ref{tab:rubric-followup-quality}) assesses how well the agent identifies and asks for withheld information before proceeding. 
\subsection{Results}
We select 9 occupations that appear frequently in our open economic index and have a sufficient number of matched MCP servers and tasks to generate scenarios grounded in 3 pairs of each. 
For each, we sampled the possible sets of 3 (task, server) pairs, resulting in a total of almost 5000 scenarios generated by Kimi-k2.5 \citep{kimiteam2026kimik25visualagentic}. These scenarios cover 130 tasks, 100 servers, and over 1000 tools.

\begin{figure}[h]
    \centering
    \includegraphics[width=\linewidth, height=6.5cm]{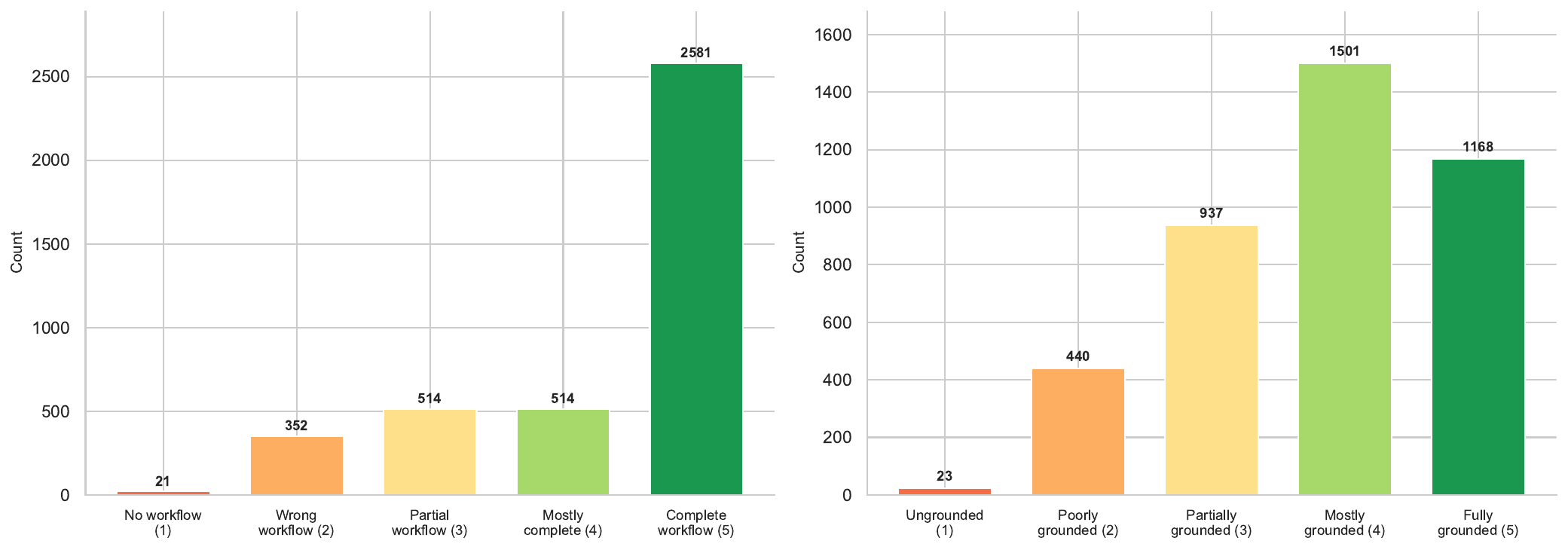}
    \caption{{Grounding and Workflow Score Counts}}
    \label{fig:hr3tscores}
\end{figure}

\begin{figure}[h!]
    \centering
    \includegraphics[width=0.75\linewidth, height=14cm]{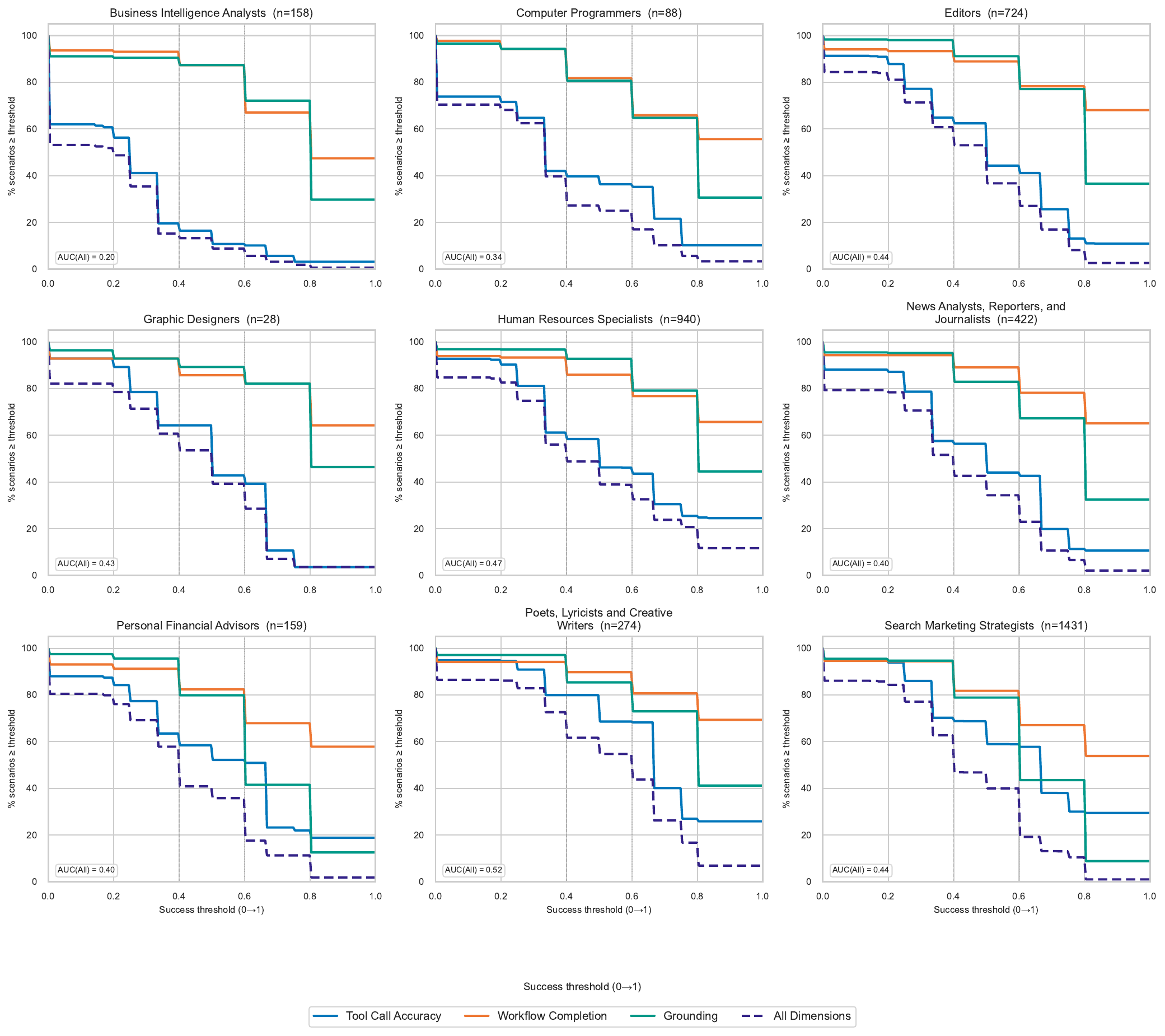}
    \caption{{Per-Occupation Score Curves}}
    \label{fig:hrv1}
\end{figure}

\textbf{Single-turn results.} We test Kimi-k2.5\footnote{ Figure \ref{fig:GPT5vKimiSmoke} shows a smoke test comparing this model choice with two OpenAI models.} with the OpenAI agent harness and virtual tools in each scenario (Kimi also acts as the tool and user). 
Figure \ref{fig:hr3tscores} provides the score distributions: about 60\% of required tool calls are executed correctly, mean workflow completion is 4.08/5, and mean grounding is 3.7/5. Per-occupation score curves, Figure \ref{fig:hrv1}, show that only a small proportion of scenarios achieve high scores across all dimensions simultaneously. We also categorize agent errors (Figure \ref{fig:hr3ttax} and Table \ref{tab:error_taxonomy}, Appendix \ref{app:error-taxonomy}), drawing on TRAIL \citep{deshpande2025trailtracereasoningagentic}.

\begin{figure}[h!]
    \centering
    \includegraphics[width=0.8\linewidth, height = 7cm]{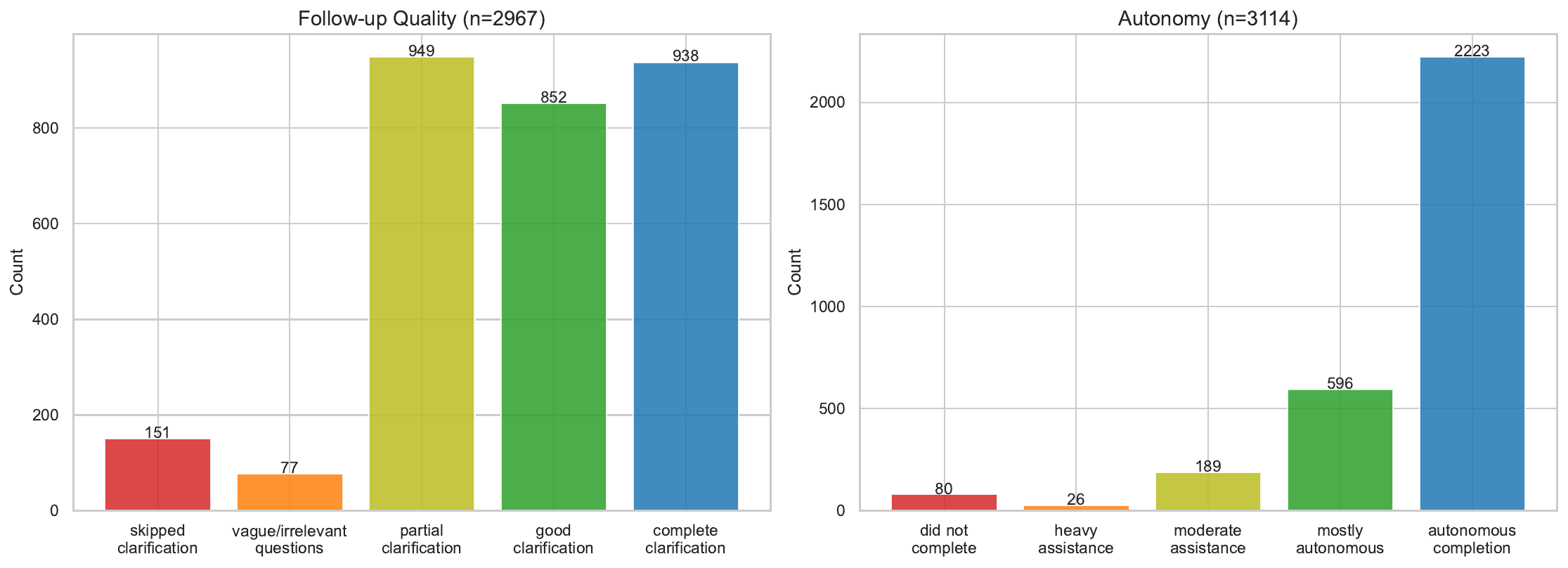}
    \caption{{Follow-up quality and autonomous score counts}}
    \label{fig:hruv1}
\end{figure}

\begin{figure}[h!]
    \centering
    \includegraphics[width=0.75\linewidth, height = 6cm]{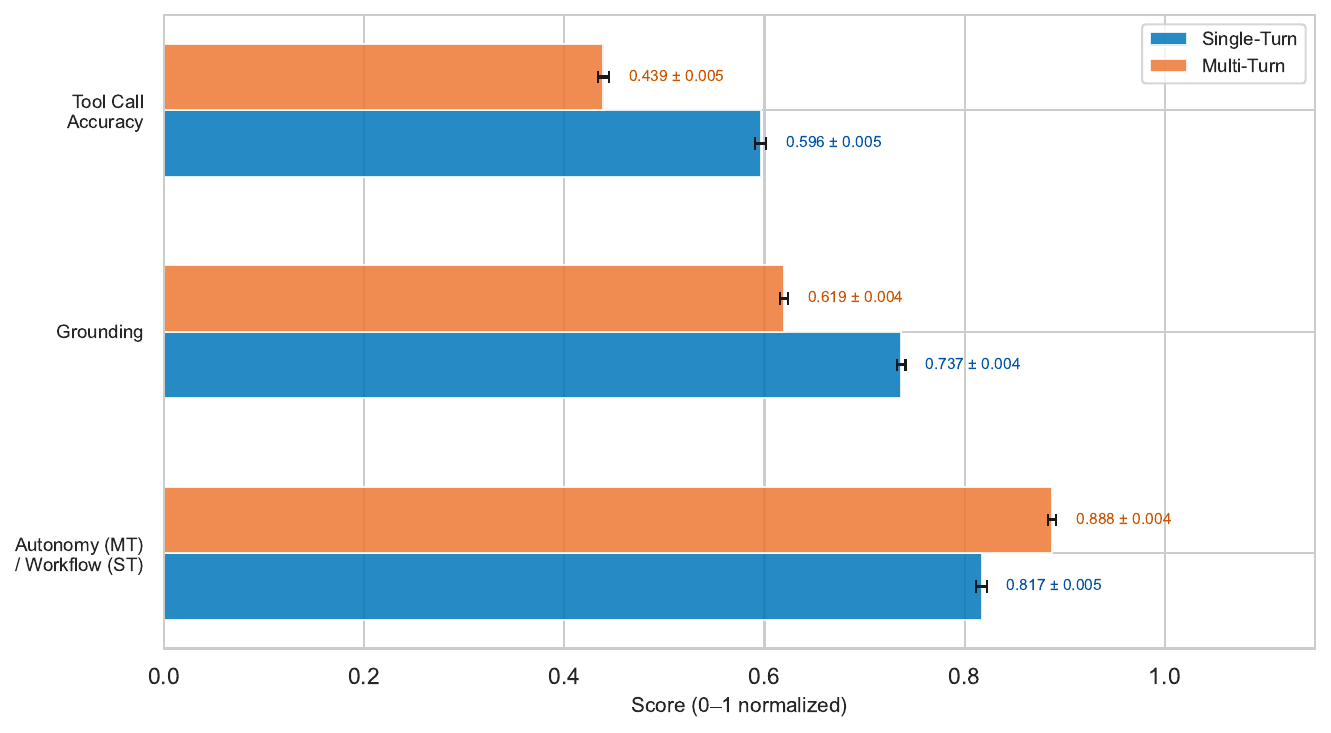}
    \caption{{Single-turn vs. Multi-turn evaluations (all scores computed over scenarios which successfully executed in both mods).}}
    \label{fig:hruv2}
\end{figure}

\textbf{Multi-turn results.} We generate information-withheld versions of the same scenarios, producing approximately 3200 valid multi-turn trajectories. Figure \ref{fig:hruv1} shows that despite mixed follow-up quality (modal outcome: partial clarification), agents often autonomously complete the workflow, suggesting they can infer enough information even with sub-par information gathering. Figure \ref{fig:hruv2} reveals that user collaboration improves workflow completion but harms tool call accuracy and grounding, suggesting that user-AI collaboration should still include granular oversight on intermediate steps.
\section{Discussion and limitations}
We have developed open source methods for measuring AI adoption and productivity across the economy. For adoption, we used the WildChat dataset and our own semantic search method to replicate the closed indexes produced by \citet{handa2025economictasksperformedai} and \citet{NBERw34255}.
To measure productivity, we developed a systematic benchmark that tests agents equipped with real or hallucinated MCP servers on scenarios grounded in O*NET tasks. The results have two main implications for the labor market. First, adoption rates are lower than the proportion of scenarios that an agent can correctly execute. Second, current agents perform best in macro level workflows when augmented by a user, though this comes with trade offs along other evaluation dimensions.  

Many open questions remain. Replacing the WildChat dataset with another collected chat dataset for more modern models would provide a more recent snapshot of adoption. Additionally, one can always enhance the available tools for the agents in our scenarios. In Appendix \ref{app: CMMC}, we analyze the coverage of O*NET software commodities by MCP servers, finding at most $33\%$ coverage. This suggests that better answers to our primary economic questions could be obtained by enhancing the server/tool list. In particular, file (pdf, slides) editing tools would allow the scenarios to incorporate a ``deliverables" aspect that is present in GDPval \citep{patwardhan2025gdpval} and RLI \citep{mazeika2025rli}.

\begin{ack}
This material is based upon work supported by the National Science Foundation under Grant Numbers 2113373 and 2414918 and a gift from OpenAI.
\end{ack}
\bibliographystyle{plainnat}
\bibliography{references, YK2, sample}

\begin{thebibliography}{43}
\providecommand{\natexlab}[1]{#1}
\providecommand{\url}[1]{\texttt{#1}}
\expandafter\ifx\csname urlstyle\endcsname\relax
  \providecommand{\doi}[1]{doi: #1}\else
  \providecommand{\doi}{doi: \begingroup \urlstyle{rm}\Url}\fi

\bibitem[Acemoglu(2021)]{acemoglu2021harms}
Daron Acemoglu.
\newblock Harms of ai.
\newblock Technical report, National Bureau of Economic Research, 2021.

\bibitem[Acemoglu(2024)]{acemoglu2024macro}
Daron Acemoglu.
\newblock The simple macroeconomics of ai.
\newblock Technical Report 32487, National Bureau of Economic Research, Cambridge, MA, May 2024.

\bibitem[Acemoglu and Restrepo(2018)]{acemoglu2018race}
Daron Acemoglu and Pascual Restrepo.
\newblock The race between man and machine: Implications of technology for growth, factor shares, and employment.
\newblock \emph{American Economic Review}, 108\penalty0 (6):\penalty0 1488--1542, 2018.

\bibitem[{AI Index Steering Committee}(2025)]{stanford2025aiindex}
{AI Index Steering Committee}.
\newblock The 2025 ai index report.
\newblock Technical report, Institute for Human-Centered AI, Stanford University, Stanford, CA, 2025.
\newblock URL \url{https://hai.stanford.edu/ai-index/2025-ai-index-report}.

\bibitem[Autor(2013)]{autor2013task}
David~H Autor.
\newblock The "task approach" to labor markets: an overview.
\newblock \emph{Journal for Labour Market Research}, 46\penalty0 (3):\penalty0 185--199, 2013.

\bibitem[Autor(2015)]{autor2015why}
David~H Autor.
\newblock Why are there still so many jobs? the history and future of workplace automation.
\newblock \emph{Journal of Economic Perspectives}, 29\penalty0 (3):\penalty0 3--30, 2015.

\bibitem[Autor et~al.(2003)Autor, Levy, and Murnane]{autor2003skill}
David~H Autor, Frank Levy, and Richard~J Murnane.
\newblock The skill content of recent technological change: An empirical exploration.
\newblock \emph{The Quarterly Journal of Economics}, 118\penalty0 (4):\penalty0 1279--1333, 2003.

\bibitem[Bick et~al.(2024)Bick, Blandin, and Deming]{bick2024rapid}
Alexander Bick, Adam Blandin, and David~J Deming.
\newblock The rapid adoption of generative ai.
\newblock Working Paper 32966, National Bureau of Economic Research, 2024.

\bibitem[Brynjolfsson et~al.(2018)Brynjolfsson, Mitchell, and Rock]{brynjolfsson2018machine}
Erik Brynjolfsson, Tom Mitchell, and Daniel Rock.
\newblock What can machines learn and what does it mean for occupations and the economy?
\newblock \emph{AEA Papers and Proceedings}, 108:\penalty0 43--47, 2018.

\bibitem[Brynjolfsson et~al.(2023)Brynjolfsson, Li, and Raymond]{brynjolfsson2023generative}
Erik Brynjolfsson, Danielle Li, and Lindsey~R Raymond.
\newblock Generative ai at work.
\newblock Technical report, National Bureau of Economic Research, 2023.

\bibitem[Chatterji et~al.(2025)Chatterji, Cunningham, Deming, Hitzig, Ong, Shan, and Wadman]{NBERw34255}
Aaron Chatterji, Thomas Cunningham, David~J Deming, Zoe Hitzig, Christopher Ong, Carl~Yan Shan, and Kevin Wadman.
\newblock How people use chatgpt.
\newblock Working Paper 34255, National Bureau of Economic Research, September 2025.
\newblock URL \url{http://www.nber.org/papers/w34255}.

\bibitem[Choi and Schwarcz(2023)]{choi2023ai}
Jonathan~H Choi and Daniel Schwarcz.
\newblock Ai assistance in legal analysis: An empirical study.
\newblock \emph{SSRN Working Paper 4539836}, 2023.

\bibitem[Cui et~al.(2024)Cui, Demirer, Jaffe, Musolff, Peng, and Salz]{cui2024effects}
Z.~K. Cui, M.~Demirer, S.~Jaffe, L.~Musolff, S.~Peng, and T.~Salz.
\newblock The effects of generative ai on high skilled work: Evidence from three field experiments with software developers.
\newblock \emph{SSRN Working Paper 4945566}, 2024.

\bibitem[Dell'Acqua et~al.(2023)Dell'Acqua, McFowland~III, Mollick, Lifshitz-Assaf, Kellogg, Rajendran, Krayer, Candelon, and Lakhani]{dellacqua2023navigating}
Fabrizio Dell'Acqua, Edward McFowland~III, Ethan~R Mollick, Hila Lifshitz-Assaf, Katherine Kellogg, Saran Rajendran, Lisa Krayer, Fran{\c{c}}ois Candelon, and Karim~R Lakhani.
\newblock Navigating the jagged technological frontier: Field experimental evidence of the effects of ai on knowledge worker productivity and quality.
\newblock Working Paper 24-013, Harvard Business School Technology \& Operations Mgt. Unit, 2023.

\bibitem[Deshpande et~al.(2025)Deshpande, Gangal, Mehta, Krishnan, Kannappan, and Qian]{deshpande2025trailtracereasoningagentic}
Darshan Deshpande, Varun Gangal, Hersh Mehta, Jitin Krishnan, Anand Kannappan, and Rebecca Qian.
\newblock Trail: Trace reasoning and agentic issue localization, 2025.
\newblock URL \url{https://arxiv.org/abs/2505.08638}.

\bibitem[Felten et~al.(2023)Felten, Raj, and Seamans]{felten2023chatgpt}
Edward Felten, Manav Raj, and Robert Seamans.
\newblock How will language modelers like chatgpt affect occupations and industries?
\newblock \emph{arXiv preprint arXiv:2303.01157}, 2023.

\bibitem[Guo et~al.(2025)Guo, Cheng, Wang, Liang, Qin, Li, Liu, Sun, and Liu]{guo2025stabletoolbenchstablelargescalebenchmarking}
Zhicheng Guo, Sijie Cheng, Hao Wang, Shihao Liang, Yujia Qin, Peng Li, Zhiyuan Liu, Maosong Sun, and Yang Liu.
\newblock Stabletoolbench: Towards stable large-scale benchmarking on tool learning of large language models, 2025.
\newblock URL \url{https://arxiv.org/abs/2403.07714}.

\bibitem[Handa et~al.(2025)Handa, Tamkin, McCain, Huang, Durmus, Heck, Mueller, Hong, Ritchie, Belonax, Troy, Amodei, Kaplan, Clark, and Ganguli]{handa2025economictasksperformedai}
Kunal Handa, Alex Tamkin, Miles McCain, Saffron Huang, Esin Durmus, Sarah Heck, Jared Mueller, Jerry Hong, Stuart Ritchie, Tim Belonax, Kevin~K. Troy, Dario Amodei, Jared Kaplan, Jack Clark, and Deep Ganguli.
\newblock Which economic tasks are performed with ai? evidence from millions of claude conversations, 2025.
\newblock URL \url{https://arxiv.org/abs/2503.04761}.

\bibitem[Hering(2023)]{hering2023indeed}
A.~Hering.
\newblock Indeed's ai at work report: How genai will impact jobs and the skills needed to perform them.
\newblock Research report, Indeed Hiring Lab, 2023.

\bibitem[Humlum and Vestergaard(2024)]{humlum2024adoption}
Anders Humlum and Emilie Vestergaard.
\newblock The adoption of chatgpt.
\newblock Technical report, Becker Friedman Institute for Economics, University of Chicago, April 2024.

\bibitem[Ling and Imas(2025)]{ling2025underreporting}
Yier Ling and Alex Imas.
\newblock Underreporting of ai use: The role of social desirability bias.
\newblock \emph{SSRN}, May 2025.
\newblock \doi{10.2139/ssrn.5232910}.
\newblock URL \url{https://ssrn.com/abstract=5232910}.

\bibitem[Liu et~al.(2024{\natexlab{a}})Liu, Yu, Zhang, Xu, Lei, Lai, Gu, Ding, Men, Yang, Zhang, Deng, Zeng, Du, Zhang, Shen, Zhang, Su, Sun, Huang, Dong, and Tang]{liu2024agentbench}
Xiao Liu, Hao Yu, Hanchen Zhang, Yifan Xu, Xuanyu Lei, Hanyu Lai, Yu~Gu, Hangliang Ding, Kaiwen Men, Kejuan Yang, Shudan Zhang, Xiang Deng, Aohan Zeng, Zhengxiao Du, Chenhui Zhang, Sheng Shen, Tianjun Zhang, Yu~Su, Huan Sun, Minlie Huang, Yuxiao Dong, and Jie Tang.
\newblock Agentbench: Evaluating {LLM}s as agents.
\newblock In \emph{The Twelfth International Conference on Learning Representations}, 2024{\natexlab{a}}.
\newblock URL \url{https://openreview.net/forum?id=zAdUB0aCTQ}.

\bibitem[Liu et~al.(2024{\natexlab{b}})Liu, Hoang, Zhang, Zhu, Lan, Kokane, Tan, Yao, Liu, Feng, Murthy, Yang, Savarese, Niebles, Wang, Heinecke, and Xiong]{liu2024apigenautomatedpipelinegenerating}
Zuxin Liu, Thai Hoang, Jianguo Zhang, Ming Zhu, Tian Lan, Shirley Kokane, Juntao Tan, Weiran Yao, Zhiwei Liu, Yihao Feng, Rithesh Murthy, Liangwei Yang, Silvio Savarese, Juan~Carlos Niebles, Huan Wang, Shelby Heinecke, and Caiming Xiong.
\newblock Apigen: Automated pipeline for generating verifiable and diverse function-calling datasets, 2024{\natexlab{b}}.
\newblock URL \url{https://arxiv.org/abs/2406.18518}.

\bibitem[Mazeika et~al.(2025)Mazeika, Gatti, Menghini, Sehwag, Singhal, Orlovskiy, Basart, Sharma, Peskoff, Lau, et~al.]{mazeika2025rli}
Mantas Mazeika, Alice Gatti, Cristina Menghini, Udari~Madhushani Sehwag, Shivam Singhal, Yury Orlovskiy, Steven Basart, Manasi Sharma, Denis Peskoff, Elaine Lau, et~al.
\newblock Remote labor index: Measuring ai automation of remote work.
\newblock \emph{arXiv preprint arXiv:2510.26787}, 2025.

\bibitem[Merali(2024)]{merali2024scaling}
Arsh Merali.
\newblock Scaling laws for economic productivity: Experimental evidence in llm-assisted translation.
\newblock \emph{arXiv preprint arXiv:2409.02391}, 2024.

\bibitem[{National Center for O*NET Development}(2025)]{onet2025online}
{National Center for O*NET Development}.
\newblock O*net online, 2025.
\newblock URL \url{https://www.onetonline.org/}.

\bibitem[Noy and Zhang(2023)]{noy2023experimental}
Shakked Noy and Whitney Zhang.
\newblock Experimental evidence on the productivity effects of generative artificial intelligence.
\newblock \emph{Science}, 381\penalty0 (6654):\penalty0 187--192, 2023.

\bibitem[OpenAI(2025)]{openai2025gptoss120bgptoss20bmodel}
OpenAI.
\newblock gpt-oss-120b and gpt-oss-20b model card, 2025.
\newblock URL \url{https://arxiv.org/abs/2508.10925}.

\bibitem[Patil et~al.(2025)Patil, Mao, Ji, Yan, Suresh, Stoica, and Gonzalez]{patil2025bfcl}
Shishir~G. Patil, Huanzhi Mao, Charlie Cheng-Jie Ji, Fanjia Yan, Vishnu Suresh, Ion Stoica, and Joseph~E. Gonzalez.
\newblock The berkeley function calling leaderboard (bfcl): From tool use to agentic evaluation of large language models.
\newblock In \emph{International Conference on Machine Learning (ICML)}, 2025.

\bibitem[Patwardhan et~al.(2025)Patwardhan, Dias, Proehl, Kim, Wang, Watkins, Fishman, Aljubeh, Thacker, Fauconnet, et~al.]{patwardhan2025gdpval}
Tejal Patwardhan, Rachel Dias, Elizabeth Proehl, Grace Kim, Michele Wang, Olivia Watkins, Sim{\'o}n~Posada Fishman, Marwan Aljubeh, Phoebe Thacker, Laurance Fauconnet, et~al.
\newblock Gdpval: Evaluating ai model performance on real-world economically valuable tasks.
\newblock \emph{arXiv preprint arXiv:2510.04374}, 2025.

\bibitem[Peng et~al.(2023)Peng, Kalliamvakou, Cihon, and Demirer]{peng2023impact}
Sida Peng, Eirini Kalliamvakou, Peter Cihon, and Mert Demirer.
\newblock The impact of ai on developer productivity: Evidence from github copilot.
\newblock \emph{arXiv preprint arXiv:2302.06590}, 2023.

\bibitem[Qin et~al.(2024)Qin, Liang, Ye, Zhu, Yan, Lu, Lin, Cong, Tang, Qian, et~al.]{qin2024toolllm}
Yujia Qin, Shihao Liang, Yining Ye, Kunlun Zhu, Lan Yan, Yaxi Lu, Yankai Lin, Xin Cong, Xiangru Tang, Bill Qian, et~al.
\newblock Toolllm: Facilitating large language models to master 16000+ real-world apis.
\newblock In \emph{International Conference on Learning Representations (ICLR)}, 2024.

\bibitem[Team(2026)]{kimiteam2026kimik25visualagentic}
Kimi Team.
\newblock Kimi k2.5: Visual agentic intelligence, 2026.
\newblock URL \url{https://arxiv.org/abs/2602.02276}.

\bibitem[Tomlinson et~al.(2025)Tomlinson, Jaffe, Wang, Counts, and Suri]{tomlinson2025working}
Kiran Tomlinson, Sonia Jaffe, Will Wang, Scott Counts, and Siddharth Suri.
\newblock Working with ai: Measuring the occupational implications of generative ai, 2025.

\bibitem[Trammell and Korinek(2023)]{trammell2023growth}
Philip Trammell and Anton Korinek.
\newblock Economic growth under transformative ai.
\newblock Working Paper 31815, National Bureau of Economic Research, October 2023.
\newblock URL \url{http://www.nber.org/papers/w31815}.

\bibitem[{U.S. Bureau of Labor Statistics}(2026)]{usbls}
{U.S. Bureau of Labor Statistics}.
\newblock Occupational employment and wage statistics, 2026.
\newblock URL \url{https://data.bls.gov/oes/}.
\newblock Accessed: 2026-02-25.

\bibitem[Wang et~al.(2025)Wang, Chang, Patel, Biju, Wu, Liu, Ding, Rezazadeh, Shah, Bao, and Siow]{wang2025mcpbench}
Zhenting Wang, Qi~Chang, Hemani Patel, Shashank Biju, Cheng-En Wu, Quan Liu, Aolin Ding, Alireza Rezazadeh, Ankit Shah, Yujia Bao, and Eugene Siow.
\newblock Mcp-bench: Benchmarking tool-using llm agents with complex real-world tasks via mcp servers.
\newblock \emph{arXiv preprint arXiv:2508.20453}, 2025.

\bibitem[Webb(2019)]{webb2019impact}
Michael Webb.
\newblock The impact of artificial intelligence on the labor market.
\newblock \emph{SSRN Working Paper}, 2019.

\bibitem[Wiles et~al.(2024)Wiles, Krayer, Abbadi, Awasthi, Kennedy, Mishkin, Sack, and Candelon]{wiles2024genai}
E.~Wiles, L.~Krayer, M.~Abbadi, U.~Awasthi, R.~Kennedy, P.~Mishkin, D.~Sack, and F.~Candelon.
\newblock Genai as an exoskeleton: Experimental evidence on knowledge workers using genai on new skills.
\newblock \emph{SSRN Working Paper 4944588}, 2024.

\bibitem[Xu et~al.(2025)Xu, Soria, Tan, Roy, Agrawal, Poovendran, and Panda]{xu2025toucansynthesizing15mtoolagentic}
Zhangchen Xu, Adriana~Meza Soria, Shawn Tan, Anurag Roy, Ashish~Sunil Agrawal, Radha Poovendran, and Rameswar Panda.
\newblock Toucan: Synthesizing 1.5m tool-agentic data from real-world mcp environments, 2025.
\newblock URL \url{https://arxiv.org/abs/2510.01179}.

\bibitem[Yao et~al.(2025)Yao, Shinn, Razavi, and Narasimhan]{yao2025taubench}
Shunyu Yao, Noah Shinn, Pedram Razavi, and Karthik Narasimhan.
\newblock $\tau$-bench: A benchmark for tool-agent-user interaction in real-world domains.
\newblock In \emph{International Conference on Learning Representations (ICLR)}, 2025.

\bibitem[Zhang et~al.(2025)Zhang, Li, Long, Zhang, Lin, Yang, Xie, Yang, Liu, Lin, Huang, and Zhou]{zhang2025qwen3embeddingadvancingtext}
Yanzhao Zhang, Mingxin Li, Dingkun Long, Xin Zhang, Huan Lin, Baosong Yang, Pengjun Xie, An~Yang, Dayiheng Liu, Junyang Lin, Fei Huang, and Jingren Zhou.
\newblock Qwen3 embedding: Advancing text embedding and reranking through foundation models, 2025.
\newblock URL \url{https://arxiv.org/abs/2506.05176}.

\bibitem[Zhao et~al.(2024)Zhao, Ren, Hessel, Cardie, Choi, and Deng]{zhao2024wildchat}
Wenting Zhao, Xiang Ren, Jack Hessel, Claire Cardie, Yejin Choi, and Yuntian Deng.
\newblock Wildchat: 1m chat{GPT} interaction logs in the wild.
\newblock In \emph{The Twelfth International Conference on Learning Representations}, 2024.
\newblock URL \url{https://openreview.net/forum?id=Bl8u7ZRlbM}.

\end{thebibliography}

\appendix

\section{Smithery MCP Server Analysis}
\label{app: servers}

Our pipeline for setting up servers is as follows.
\begin{enumerate}
    \item We crawl smithery and produce an LLM generated analysis of 1520 servers. Additionally, we attempt to connect to each server and list its tools, producing a total of 627 servers that can actually be connected to by an agent without additional enterprise purchases or specific API keys.
    \item Next, for each of the 627 validated servers, we have an LLM attempt to call each tool with realistic inputs and evaluate if the response returns "real, non-error data consistent with its described purpose." Of the 627 validated servers, 416 had at least one tool of quality. In aggregate, about 33\% produced quality outputs. 
\end{enumerate}
Tables \ref{tab:server-quality} and \ref{tab:agent-errors} break down the results further. The pipeline was robust with few agent errors, the majority of which were mcp servers exceeding the context length of kimi-k2. Our analysis includes an additional check that has not yet been discussed: we check servers for ``statefulness". A server is defined as stateful if it "has tools that allow an agent to store information, modify existing data, or change the internal state of the system". It is stateless if "it provides read-only access; allows fetching, searching, or processing data without permanently altering the system's state or storing new information across calls". The intent here is to measure the complexity of the tools in another dimension; stateful tools are more realistic for what an autonomous AI needs to utilize. We see that stateless tools outnumber stateful tools at a 2-to-1 ratio.

\begin{table}[h]
\centering
\caption{MCP Server Quality Statistics (Smithery, February 2025)}
\label{tab:server-quality}
\begin{tabular}{lr}
\toprule
\textbf{Metric} & \textbf{Count} \\
\midrule
Total servers crawled                        & 1,520 \\
Servers passing validation                   & 627   \\
\addlinespace
Servers with $\geq$1 passing tool            & 416   \\
\quad Stateful                               & 131   \\
\quad Stateless                              & 285   \\
\addlinespace
Total tools passing quality check            & 2,763 \\
\addlinespace
Total tools called            & 8,233 \\
\bottomrule
\end{tabular}
\end{table}

\begin{table}[h]
\centering
\caption{Agent Errors Encountered During Server Quality Checking}
\label{tab:agent-errors}
\begin{tabular}{lrr}
\toprule
\textbf{Error Type} & \textbf{Count} & \textbf{\% of Valid Servers} \\
\midrule
Context length exceeded   & 12 & 1.9\% \\
Invalid MCP response      &  5 & 0.8\% \\
\midrule
Total agent errors        & 17 & 2.7\% \\
\bottomrule
\end{tabular}
\end{table}

\subsection{O*NET commodity coverage}
\label{app: CMMC}
As a test of the ``breadth" covered by MCP servers, we measure how well they cover the 137 O*NET technology skill "commodities" (standardized software categories like "Accounting software", "CRM software", etc.). Our methodology remains the same: use Qwen3-Embedding-8B to embed both commodity definitions and server/tool descriptions, then match via cosine similarity (threshold 0.5). For example, a commodity embedding may resemble 
\textit{Instruct: Retrieve software tools that provide similar functionality as software in the following category.
Query: Software to enable business functions related to accounting and management}
For more generous retrieval, we also try querying each commodity example, and matching if any examples exceed the similarity score threshold.

\textit{Instruct: Retrieve software tools that provide similar functionality...
Query: Software to enable business functions relating to accounting and management such as QuickBooks}

\textit{Instruct: Retrieve software tools that provide similar functionality...
Query: Software to enable business functions relating to accounting and management such as Sage 50}

The results for the validated MCP servers are given below.

\begin{table}[H]
\centering
\caption{Commodity--Server Matching Coverage (Smithery MCP Servers)}
\label{tab:smithery-coverage}
\begin{tabular}{lcc}
\toprule
Configuration & Commodity Coverage & Server Utilization \\
\midrule
Baseline                         & 21 / 137 (15.3\%) & 58 / 1356 (4.3\%) \\
\texttt{--include-commodity-examples}      & 65 / 137 (47.4\%) & 330 / 1356 (24.3\%) \\
\texttt{--validated-only}                  & 13 / 137 (9.5\%)  & 17 / 555 (3.1\%) \\
Both flags                       & 49 / 137 (35.8\%) & 90 / 555 (16.2\%) \\
\bottomrule
\end{tabular}
\end{table}
\subsection{Optional Tool Simulation}
At scale, Smithery MCP servers are not free to call (it is free up to a certain usage). Given this, we also implement a virtual tool system that can ingest MCP server documentation and produce realistic tool outputs (using an LLM). 
\begin{center}
\begin{tikzpicture}[
    node distance=0.8cm and 1.5cm,
    >={Stealth[length=2mm]},
    process/.style={
        rectangle, rounded corners=3pt, draw=#1, fill=#1!12,
        text width=3.8cm, minimum height=0.7cm,
        align=center, font=\sffamily\scriptsize, line width=0.7pt
    },
    decision/.style={
        diamond, draw=validgold, fill=validgold!15,
        text width=1.6cm, minimum height=0.6cm,
        align=center, font=\sffamily\scriptsize, line width=0.7pt,
        inner sep=1pt, aspect=2.2
    },
    result/.style={
        rectangle, rounded corners=3pt, draw=#1, fill=#1!15,
        text width=3.2cm, minimum height=0.6cm,
        align=center, font=\sffamily\scriptsize, line width=0.7pt
    },
    infobox/.style={
        rectangle, rounded corners=2pt, draw=#1!60, fill=#1!8,
        text width=3.6cm, align=left,
        font=\sffamily\tiny, inner sep=4pt, line width=0.5pt
    },
    arrow/.style={->, line width=0.7pt, color=black!70},
    dasharrow/.style={->, line width=0.5pt, dashed, color=black!50},
    lbl/.style={font=\sffamily\tiny, color=black!60},
]


\node[process=agentblue] (agent)
    {\textbf{Agent Tool Call}\\[1pt]
     \texttt{tool\_name}, \texttt{arguments}};

\node[process=agentblue, below=of agent] (vtool)
    {\textbf{Virtual Tool Function}\\[1pt]
     \texttt{dynamic\_run\_function}};

\node[process=promptgreen, below=of vtool] (assembly)
    {\textbf{Prompt Assembly}\\[1pt]
     \texttt{build\_tool\_sim\_request}};

\node[process=llmpurple, below=1.0cm of assembly] (llm)
    {\textbf{Scenario Master LLM}\\[1pt]
     Structured JSON output};

\node[decision, below=1.0cm of llm] (valid)
    {\textbf{Valid?}};

\node[result=successgreen, below left=1.0cm and 1.0cm of valid] (success)
    {\texttt{\{"error": "",}\\
     \texttt{"response": "..."\}}};

\node[result=failred, below right=1.0cm and 1.0cm of valid] (fail)
    {\texttt{\{"error": "...",}\\
     \texttt{"response": ""\}}};

\node[process=agentblue, below=2.6cm of valid] (ret)
    {\textbf{Result Returned to Agent}\\[1pt]
     Append to shared simulation history};

\draw[arrow] (agent) -- (vtool);
\draw[arrow] (vtool) -- (assembly);
\draw[arrow] (assembly) -- (llm);
\draw[arrow] (llm) -- (valid);
\draw[arrow] (valid) -- node[lbl, left, xshift=-1pt] {Yes} (success);
\draw[arrow] (valid) -- node[lbl, right, xshift=1pt] {No} (fail);
\draw[arrow] (success) |- (ret);
\draw[arrow] (fail) |- (ret);


\node[infobox=promptgreen, right=1.8cm of assembly] (promptinfo)
    {\textbf{Simulation Request}\\[2pt]
     \textbullet\ Server: ID, name, desc.\\
     \textbullet\ Tool: name, description\\
     \textbullet\ Input schema (JSON)\\
     \textbullet\ Tool call args (JSON)};
\draw[dasharrow] (assembly.east) -- (promptinfo.west);

\node[infobox=llmpurple, right=1.8cm of llm] (llminfo)
    {\textbf{Message Sequence}\\[2pt]
     1.\ System prompt\\
     2.\ Conversation history\\
     3.\ Prior tool sim. messages\\
     4.\ Current sim. request};
\draw[dasharrow] (llm.east) -- (llminfo.west);

\node[infobox=validgold, right=1.8cm of valid] (validinfo)
    {\textbf{Validation Checks}\\[2pt]
     1.\ Required args present?\\
     2.\ No hallucinated args?\\
     3.\ Types match schema?};
\draw[dasharrow] (valid.east) -- ++(0.5,0) |- (validinfo.west);

\node[above=0.3cm of agent, font=\sffamily\small\bfseries, color=black!80]
    {Virtual Tool Simulation Pipeline};

\end{tikzpicture}
\end{center}

\section{Evaluation Methodology}
\label{app:eval-methodology}

After an LLM agent runs on a benchmark scenario (producing a trajectory of tool calls and responses), the trajectory is evaluated across three complementary dimensions using LLM-as-judge.

\textbf{Tool Call Accuracy:} This dimension mimics the Berkeley Function Calling Leaderboard (BFCL) and evaluates whether the agent called the correct tools with the correct arguments. All calls from the agent's trajectory are extracted. A basic auto-scoring handles failure cases without LLM involvement: a score of 0 is automatically assigned when a scenario fails catastrophically (e.g., agent max turns exceeded) or when the agent never called a tool present in the answer key at all. Next, as is done in BFCL, we compare each tool call in the answer key to the last actual call in the trajectory of the same tool name (in the case where M calls of the same tool are expected, an array of the last M calls is compared). An LLM judge is prompted to determine functional equivalence; a correct call may match the expected arguments exactly or may be a functionally equivalent rephrasing (e.g., different capitalization, minor wording variation, same intent). The judge outputs a binary score: 1 (correct) or 0 (incorrect), along with reasoning.

\textbf{Tool Workflow Completion:} This dimension evaluates whether the agent followed the intended multi-step tool-use workflow, focusing on the macro-level organization of tool calls rather than on granular tool calls. The LLM judge is provided with the query, the expected tool workflow, a separate analysis of how each tool contributes, and a condensed agent trajectory showing reasoning, tool calls, tool outputs, and the final response. The judge evaluates: (i) Was the sequence of tools correct? (ii) Did data flow between tools correctly? (iii) Was the final response a coherent integration of the workflow results? A score between 1 and 5 is produced using the rubric given in Table \ref{tab:workflow-rubric}.

\textbf{Grounding:} This dimension evaluates whether the agent's response is supported by the tool outputs that the agent actually received, as opposed to hallucinated responses. The judge is provided with the query and the agent trajectory, including tool calls and any claims in response to the query. The judge is instructed to enumerate grounded claims (traceable to tool outputs) and ungrounded claims (asserted without evidence), then assigns an overall score based on Table \ref{tab:grounding-rubric}.

For multi-turn trajectories, two additional dimensions are evaluated. \textbf{Autonomy} measures how independently the agent completed the workflow without user steering, and \textbf{Follow-up Quality} assesses how well the agent identified and asked for deliberately withheld information before proceeding with tool calls.

\subsection{Evaluation Rubrics}
\label{app:rubrics}

\begin{table}[H]
\centering
\caption{Workflow Completion Rubric}
\label{tab:workflow-rubric}
\begin{tabular}{clp{10cm}}
\toprule
\textbf{Score} & \textbf{Label} & \textbf{Meaning} \\
\midrule
1 & No Workflow       &  The agent did not attempt the expected workflow. \\
\addlinespace
2 & Wrong Workflow    & Tools were called but in the wrong order, wrong tools were used, or outputs were incorrectly passed between steps. \\
\addlinespace
3 & Partial Workflow  & Some steps were executed correctly, but a critical step was missed. \\
\addlinespace
4 & Mostly Complete   & The correct tool sequence was followed but there are minor gaps. \\
\addlinespace
5 & Complete Workflow & The full workflow was executed: correct tool ordering, proper inter-tool data passing, and a coherent final response. \\
\bottomrule
\end{tabular}
\end{table}

\begin{table}[H]
\centering
\caption{Grounding Rubric}
\label{tab:grounding-rubric}
\begin{tabular}{clp{10cm}}
\toprule
\textbf{Score} & \textbf{Label} & \textbf{Meaning} \\
\midrule
1 & Ungrounded         & Mostly fabricated; no meaningful connection to tool outputs. \\
\addlinespace
2 & Poorly Grounded    & Some claims connect to tool outputs, but the majority are unsupported. \\
\addlinespace
3 & Partially Grounded & Key claims are grounded, but there are notable unsupported assertions. \\
\addlinespace
4 & Mostly Grounded    & The majority of claims are traceable to tool outputs; some acceptable inferences. \\
\addlinespace
5 & Fully Grounded     & All factual claims are directly supported by tool outputs. \\
\bottomrule
\end{tabular}
\end{table}

\begin{table}[H]
\centering
\caption{Autonomy Rubric}
\label{tab:rubric-autonomy}
\begin{tabular}{clp{10cm}}
\toprule
\textbf{Score} & \textbf{Label} & \textbf{Meaning} \\
\midrule
1 & Did not complete      & Workflow was not completed (including turn-expired scenarios), or required heavy user steering and still failed to finish. \\
\addlinespace
2 & Heavy assistance      & Completed little of the workflow without substantial user redirection; the user had to repeatedly steer core execution. \\
\addlinespace
3 & Moderate assistance   & Partial-to-mostly completed, but multiple meaningful user interventions were needed to correct or unblock progress. \\
\addlinespace
4 & Mostly autonomous     & Workflow completed with limited user steering; clarifications were mostly essential and execution was largely self-directed. \\
\addlinespace
5 & Autonomous completion & Workflow completed end-to-end autonomously, except for essential unknown inputs the user needed to provide. \\
\bottomrule
\end{tabular}
\end{table}

\begin{table}[H]
\centering
\caption{Follow-up Quality Rubric}
\label{tab:rubric-followup-quality}
\begin{tabular}{clp{10cm}}
\toprule
\textbf{Score} & \textbf{Label} & \textbf{Meaning} \\
\midrule
1 & Skipped clarification         & The agent proceeded with tool calls without asking for withheld parameters, using guessed or placeholder values. \\
\addlinespace
2 & Vague or irrelevant questions & The agent asked questions but they were too vague to elicit the specific withheld values, or were tangential to the missing parameters. \\
\addlinespace
3 & Partial clarification         & The agent asked for some but not all withheld parameters, or missed a key parameter requiring re-prompting. \\
\addlinespace
4 & Complete with minor issues    & The agent asked for all required withheld parameters but with some inefficiency: extra turns, awkward phrasing, or slight redundancy. \\
\addlinespace
5 & Complete clarification        & The agent identified all withheld parameters, asked clear and targeted questions, and correctly used the provided values in subsequent tool calls. \\
\bottomrule
\end{tabular}
\end{table}

\section{Scenario Examples}
\label{app:scenario-examples}
\lstdefinestyle{promptstyle}{
  basicstyle=\ttfamily\footnotesize,
  breaklines=true,
  breakatwhitespace=false,
  columns=fullflexible,
  keepspaces=true,
  frame=single,
  framesep=6pt,
  backgroundcolor=\color{gray!8},
  rulecolor=\color{gray!50},
  xleftmargin=6pt,
  xrightmargin=6pt,
  aboveskip=8pt,
  belowskip=8pt,
}

Below are examples of single-turn and multi-turn (information-withheld) benchmark scenarios.

\begin{lstlisting}[style=promptstyle, label={lst:example-scenario-v1}]
Occupation: Human Resources Specialist
O*NET Tasks:
  1. Provide management with information or training related to interviewing,
     performance appraisals..
  2. Schedule or administer skill, intelligence, psychological, or drug tests
     for current or prospective employees.
User Request:
  "I need to set up a mandatory Python skills assessment for Sarah Chen from
   Software Engineering. She has been selected for a technical skills audit
   and I need to verify her programming competency. Please search for her in
   our employee system, check her training leave balance (she'll need 1 day
   for the assessment), create a Python fundamentals question about data
   structures, build a single-question assessment quiz, and schedule the exam
   for her on February 15th, 2024 (9 AM to 5 PM window). Make sure to assign
   her to the exam and provide me with the assessment URL so I can send it to
   her manager."
Tool Analysis:
  The EduBase server provides assessment and exam management tools that align
  with HR tasks of screening and testing employees and candidates. The Leave
  Manager server tracks employee time off, supporting availability checks
  before scheduling assessments.
Cross-Tool Workflow:
  1. [Parameter dependency] Search for the employee to confirm identity and
     obtain the employee ID.
  2. [Parameter dependency] Use the employee ID to check leave balance,
     ensuring availability for testing/training days.
  3. [Parameter dependency] Create a skills assessment quiz in EduBase and
     assign it to the employee, using the confirmed employee identifier to
     grant exam access.
  4. [Conditional routing] Retrieve quiz results to determine next steps:
     if score < 80%, schedule additional training; otherwise mark complete.
\end{lstlisting}
\begin{lstlisting}[style=promptstyle, label={lst:example-scenario}]
User Request:
  "I need to move fast on filling that senior opening. Can you pull everyone
   who's shown interest, schedule them for an interview soon, reach out with
   a call and a confirmation email, and then generate the hiring paperwork
   draft once they accept?"
Withheld Information:
  - job_requisition_title: The exact job title being recruited for, used to
    filter leads/contacts. Value: "Senior Data Engineer"
  - interview_datetime: ISO datetime for the interview slot to be scheduled.
    Value: "2024-07-10T14:00:00Z"
  - hiring_manager_id: HubSpot owner ID of the recruiter who owns the meeting,
    calls, and emails. Value: "901234567"
Target Follow-Up Questions:
  1. "What job title are you actively recruiting for right now?"
  2. "When do you want to schedule the initial interview (date and time)?"
  3. "Who is the hiring manager or recruiter that should own the interview,
     calls, and emails?"
\end{lstlisting}
\section{User Implementation}
\label{app: user}
There are two critical details for implementing an agent that acts as a realistic user. The first is \emph{role inversion}; the user operates in a session where it believes it is the assistant, and agent messages arrive as messages from the user (possibly transcribed with some indication that it is the assistant responding). Second, the user agent does not see any intermediate reasoning steps or tool calls.  Figure \ref{fig:role inversion} demonstrates both of these details.

\begin{figure}[H]
\centering
\begin{minipage}{0.85\textwidth}
\begin{lstlisting}[style=conversationstyle, label={lst:role-inversion}]
        ASSISTANT SESSION               USER SESSION
        ===============               ============
        [sys] student.md              [sys] user.md + scenario

                                      [asst] "I need to set up a
                                        mandatory Python skills
                                      assessment for Sarah Chen..."
                                      (pre-filled as own message)

        [user] "I need to set up a
        mandatory Python skills
        assessment for Sarah Chen..."

        [asst] <tool_call>
        smart_employee_search(...)
        [func] { "id": 4721, ... }
        [asst] <tool_call>
        get_leave_balance(4721)
        [func] { "balance": 8, ... }
        [asst] "I found Sarah Chen    [user] "The assistant responded:
        (ID 4721). She has 8 leave    I found Sarah Chen (ID 4721).
        days remaining..."         ->  She has 8 leave days..."

                                      [asst] "Great, now create the
                                      quiz and schedule the exam."

        [user] "Great, now create
        the quiz and schedule
        the exam."

        [asst] <tool_call>
        edubase_post_question(...)
        ...
        [asst] "Done! The exam is     [user] "The assistant responded:
        scheduled for Feb 15.      ->  Done! The exam is scheduled..."
        Here is the URL: ..."
                                      [asst] "<END_CONVERSATION>"
\end{lstlisting}
\end{minipage}
\caption{Role inversion in the multi-agent loop.}
\label{fig:role inversion}
\end{figure}

\section{Additional Plots and Tables}
\label{app:addlplots}

\subsection{Economic Index}
\begin{figure}[H]
    \centering
    \includegraphics[width=1\linewidth, height = 10cm]{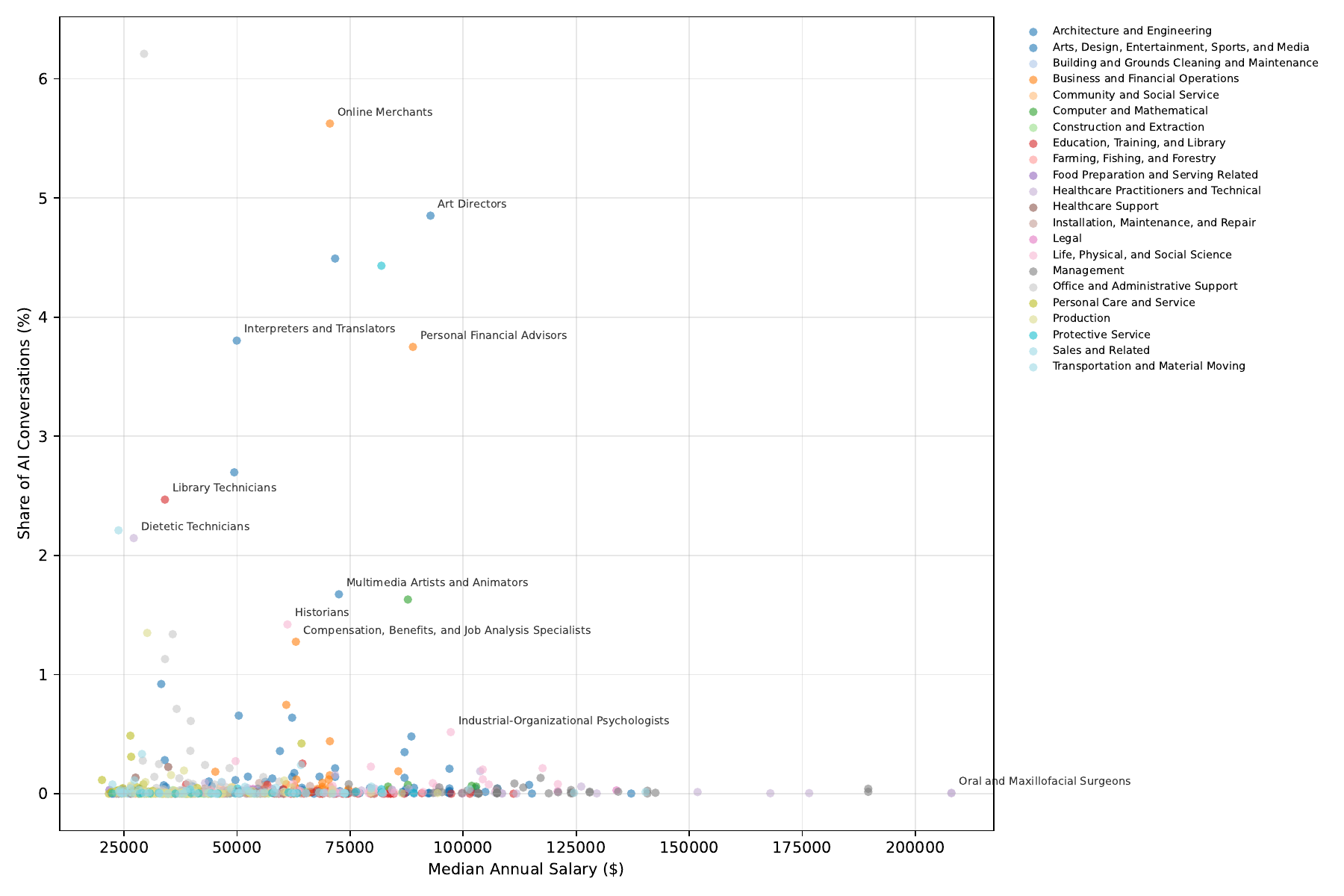}
    \caption{{AI usage vs. Median Wage. Wage data taken from Anthropic.}}
    \label{fig:OEIWagevUsage}
\end{figure}

Figure \ref{fig:OEIWagevUsage} explores the relationship between AI usage and median wage across occupations, providing additional context for the adoption patterns described in the main text.

\subsection{Benchmark Evaluation}
\begin{figure}[H]
    \centering
    \includegraphics[width=1\linewidth, height = 10cm]{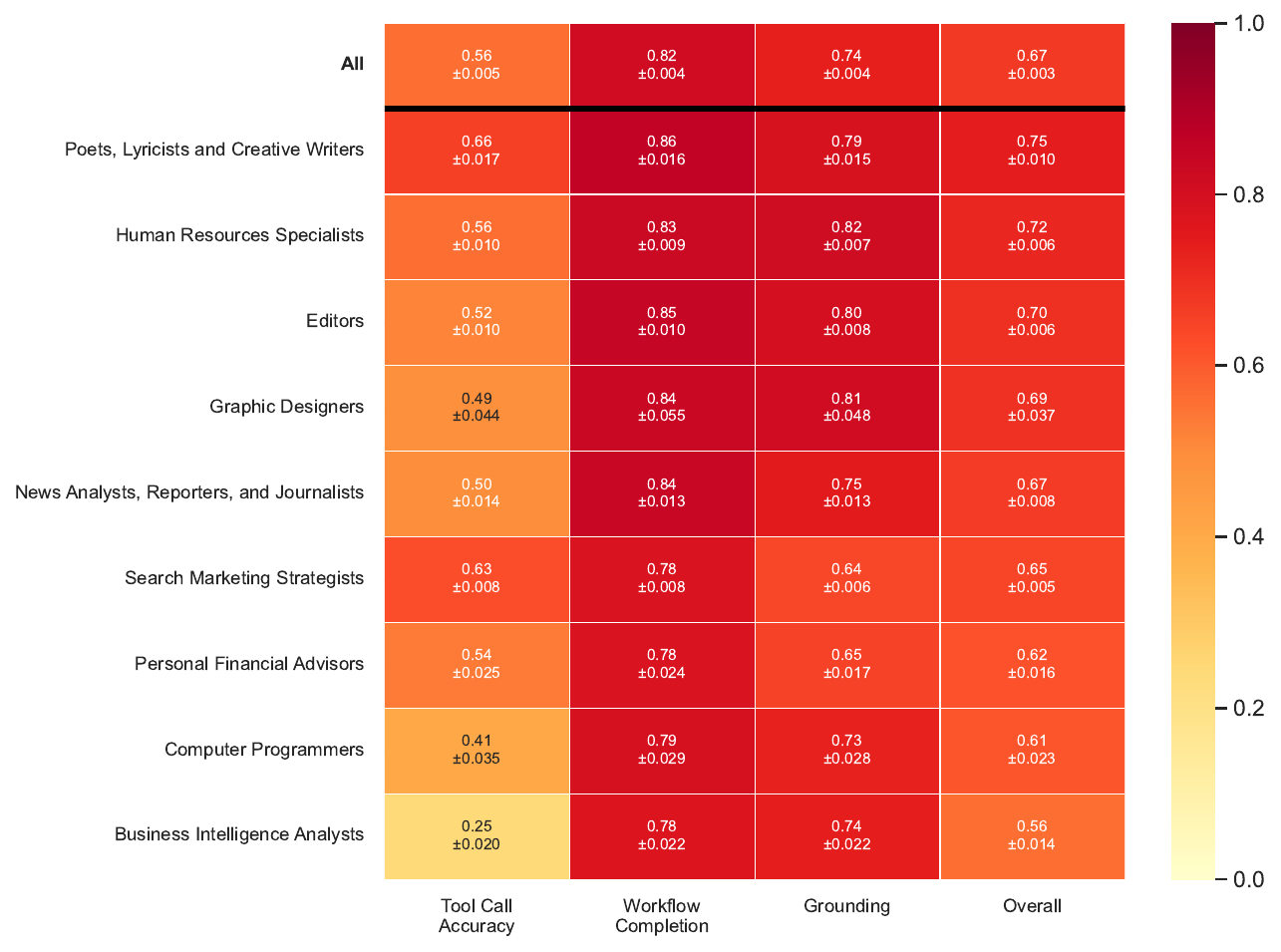}
    \caption{{Per-Occupation Performance Heatmap}}
    \label{fig:hrv0}
\end{figure}

Figures \ref{fig:hrv0}--\ref{fig:hrv1} provide supplementary views of benchmark performance across occupations and evaluation dimensions.
\begin{table}[h]
\centering
\caption{Correlations between evaluation dimensions and total tool count.}
\label{tab:tool_count_correlations}
\begin{tabular}{lccc}
\toprule
Evaluation Dimension & $\rho$ & $p$-value & Sig. \\
\midrule
  Tool Call Accuracy & $-0.2700$ & $< 0.001$ & *** \\
  Workflow Completion & $-0.0453$ & $0.003$ & ** \\
  Grounding & $+0.0120$ & $0.436$ &  \\
\bottomrule
\end{tabular}
\end{table}
\begin{figure}[H]
    \centering
    \includegraphics[width=1\linewidth, height = 7cm]{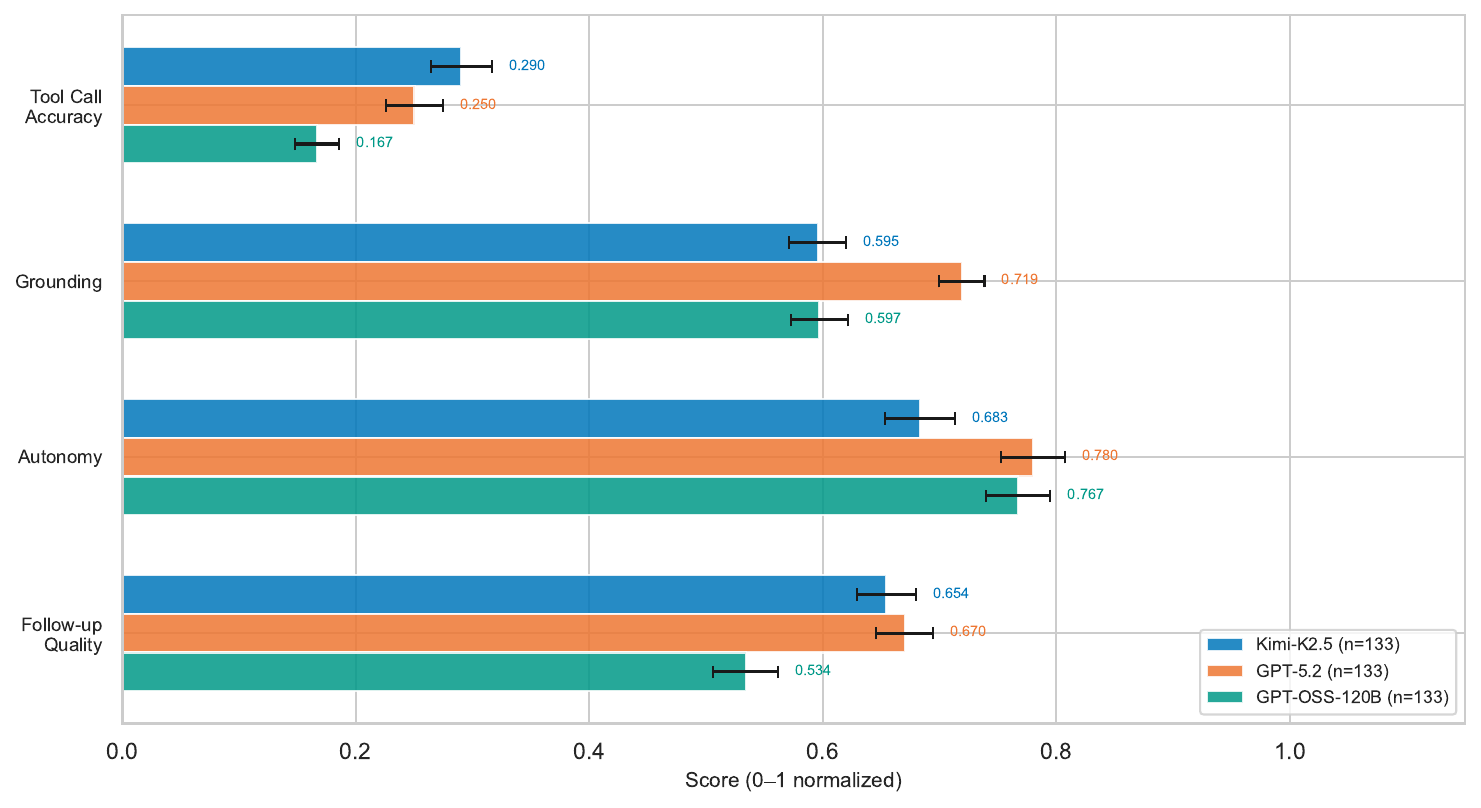}
    \caption{{GPT-5.2 vs. Kimi-k2.5 Smoke Test}}
    \label{fig:GPT5vKimiSmoke}
\end{figure}

\section{Agentic Error Taxonomy}
\label{app:error-taxonomy}

\begin{figure}[H]
    \centering
    \includegraphics[width=0.75\linewidth, height = 12cm]{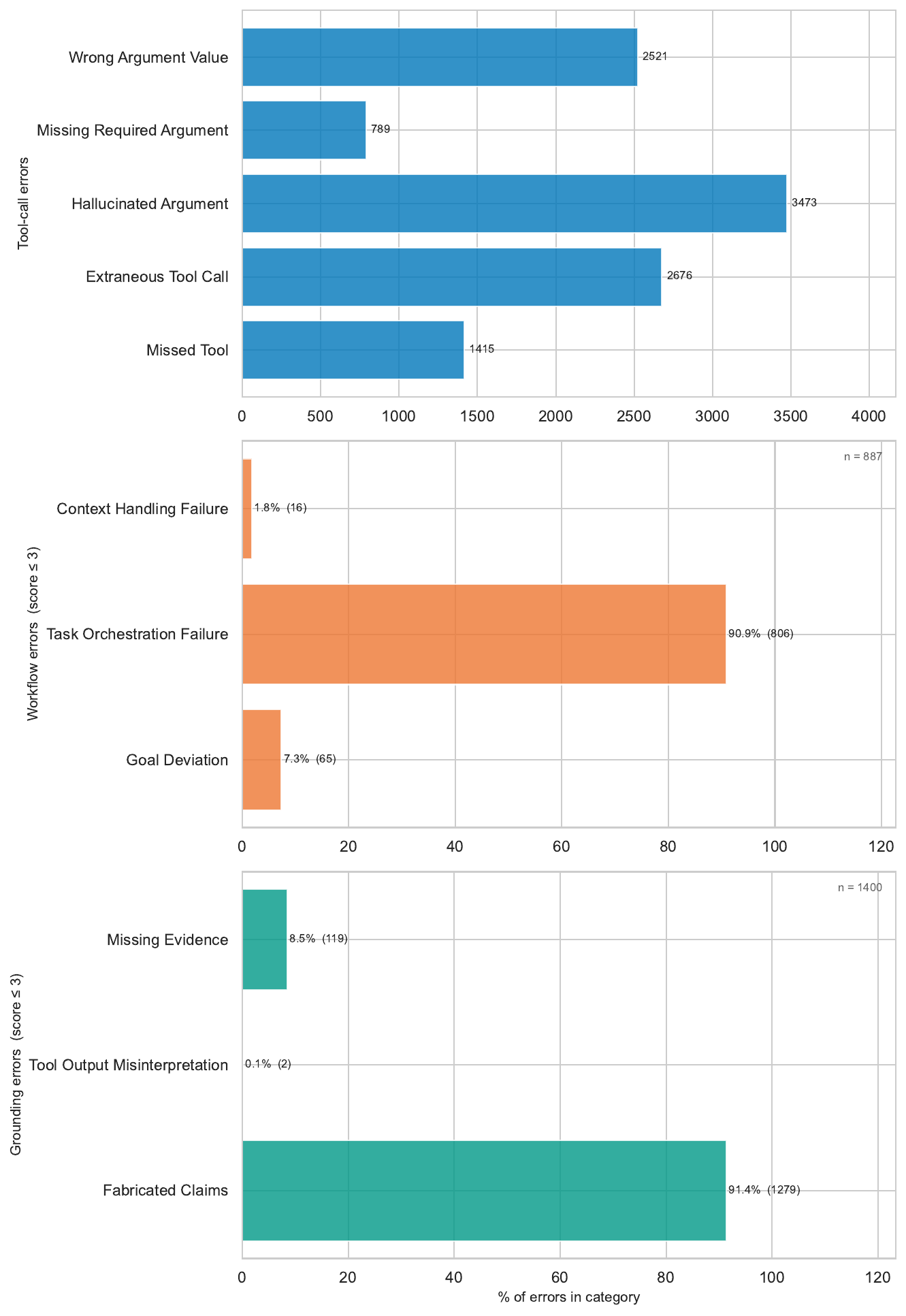}
    \caption{{Agentic Error Taxonomy.
    }}
    \label{fig:hr3ttax}
\end{figure}

We categorize agent errors across trajectories using reasoning outputs from the evaluators and gpt-oss-120b. \textit{Tool-call errors} combine three sources: extraneous tool calls, missed tools in which a required tool was never called (identified by automated scoring), and argument-level errors in which the tool was called but with incorrect arguments (LLM-judged calls with a score of 0, classified into three categories: hallucinated argument, missing required argument, and wrong argument value). \textit{Workflow} and \textit{grounding} errors are derived from evaluator reasoning strings for agentic trajectories scoring below three. Along the workflow dimension, we categorize into Goal Deviation (``agents distracted from primary objectives''), Task Orchestration Failure (``failure in the ordering of tasks executed''), and Context Handling Failure (``agent inability to pass data from one task to the next''). We derive several of these categories from TRAIL \citep{deshpande2025trailtracereasoningagentic}. Table \ref{tab:error_taxonomy} provides full definitions; Figure \ref{fig:hr3ttax} shows the frequency distribution of each error type.

\begin{table}[h]
\centering
\caption{Error taxonomy categories used for classifying agent failures, with definitions and anchoring to the TRAIL taxonomy \citep{deshpande2025trailtracereasoningagentic}.}
\label{tab:error_taxonomy}
\tiny
\begin{tabular}[]{p{1.5cm}p{2.8cm}p{5cm}p{5cm}}
\toprule
Dimension & Category & Description & TRAIL Anchor \\
\midrule
  Tool Call & Missed Tool & An expected tool was never called by the agent. & Tool Selection Errors \\
  Tool Call & Extraneous Tool & An unnecessary tool was called. & Reasoning errors -- poor info. retrieval \\
  Tool Call & Hallucinated Argument & Argument name absent from schema. & --- \\
  Tool Call & Missing Required Argument & A mandatory parameter is omitted from the tool call. & --- \\
  Tool Call & Wrong Argument Value & Argument present incorrect. & --- \\
\midrule
  Workflow & Goal Deviation & Agent pursues a different objective than specified. & Goal Dev. / Inst. non-comp. / Incorrect ID. \\
  Workflow & Task Orchestration Failure & Incorrect step sequencing, skipped dependencies, or violated ordering. & Task management -- task orchestration \\
  Workflow & Context Handling Failure & Information lost, ignored, or not carried forward. & Context management -- context handling failures \\
\midrule
  Grounding & Fabricated Claims & Statements with no basis in any tool output or provided context. & Hallucinations \\
  Grounding & Tool Output Misinterpretation & Tool output retrieved but misread, misrepresented, or incorrectly applied. & Reasoning errors -- tool output misrepresentation \\
  Grounding & Missing Evidence & Claim made without retrieving or citing accessible tool output. & --- \\
\bottomrule
\end{tabular}
\end{table}

\section{Related Work}
\label{app:related-work}

\textbf{Task-centric automation analysis:}
The idea of studying labor through discrete tasks is not new. \citet{autor2003skill} argue that computer capital can replace or augment labor capital depending on the complexity of the discrete task; \citet{autor2013task} posits that tasks can illuminate the interaction between wage growth and the available labor supply, offshoring, and technology. Other work utilizes the task-centric lens to argue that automation \emph{increases} labor demand by creating new tasks or augmenting labor on existing tasks \citep{autor2015why, acemoglu2018race}.

\textbf{Measuring and Forecasting AI Usage:} Our work also builds on prior efforts to measure how AI is currently being used and to forecast its future impact on the economy. Empirical studies on AI usage for \emph{singular} occupations have been conducted for  writing \citep{noy2023experimental}, customer service \citep{brynjolfsson2023generative}, software engineering \citep{peng2023impact, cui2024effects}, translation \citep{merali2024scaling}, data science \citep{wiles2024genai}, legal analysis \citep{choi2023ai}, and consulting \citep{dellacqua2023navigating}. \citet{humlum2024adoption} and \citet{bick2024rapid} conduct surveys to measure AI usage in the workforce. Forecasting efforts for automation through computerization \citep{webb2019impact} and traditional machine learning \citep{brynjolfsson2018machine} involve using human annotation to label the automation potential of O*NET occupations and work activities. By linking AI applications with O*NET abilities \citet{felten2023chatgpt}, estimate that law and finance are the industries most exposed to LLM automation. \citet{merali2024scaling} forecast LLM productivity in translation using scaling laws. \citet{acemoglu2024macro} and \citet{trammell2023growth} provide a theoretical macroeconomic framework to forecast the potential productivity gains and GDP growth from AI. Indeed hiring data is used by \citet{hering2023indeed} to forecast the impact of AI on the demand for various labor skills. \citet{handa2025economictasksperformedai}, \citet{tomlinson2025working}, and \citet{NBERw34255} pioneered the idea of mapping chat data to O*NET objects to produce a snapshot of AI usage.

\textbf{Benchmarking LLM tool usage and agentic capabilities:}
We test AI on O*NET tasks by assessing tool usage. The Berkeley Function Calling Leaderboard (BFCL) is the standard for measuring technical agentic tool calling accuracy via a strict abstract search tree method \citep{patil2025bfcl}. Tools on our MCP servers often have arguments that require ``natural language'' inputs. Given this, we differ and use LLM-as-a-judge to evaluate tool calls. Several other tool-calling datasets and benchmarks have been introduced. Using StableToolBench \citep{guo2025stabletoolbenchstablelargescalebenchmarking}, \citet{liu2024apigenautomatedpipelinegenerating} create a series of verifiable tool calling questions and answers. With the RapidAPIHub, \citet{qin2024toolllm} introduced tool-bench, an instruction-tuning dataset and benchmark for tool usage. \citet{xu2025toucansynthesizing15mtoolagentic} introduces the practice of designing tool calling simulations by connecting agents to MCP servers hosted on Smithery, but does not attempt thorough evaluations of the trajectories. \citet{wang2025mcpbench} adds this, building a systematic benchmark to test models on large scale tasks that require the use of 28 possible MCP servers. Our systematic economic benchmark builds on this by grounding the tasks in O*NET data, utilizing 416 MCP servers. $\tau$-bench presents a more economically grounded evaluation by testing agents in domain specific scenarios (Retail and Airline) where they must interact with a simulated user \citep{yao2025taubench}. \emph{Agentbench} adds realism by incorporating a simulated user into agentic benchmarks \citep{liu2024agentbench}.


\section{Prompts}
\label{app:prompts}

This appendix contains the full text of all LLM prompts used in data generation and evaluation. Template placeholders appear in curly braces (e.g., \texttt{\{OCCUPATION\}}) and are filled programmatically at runtime. We remark that Claude-Opus-4.6 assisted in prompt generation.

\subsection{Chat occupational relevancy classification}
This prompt instructs the model to classify a user-LLM conversation from WildChat as occupationally relevant or irrelevant. 
\begin{lstlisting}[style=promptstyle]
Is the user's request occupationally relevant (related to work, professional tasks, or career activities)? Respond in exactly this format:
[Yes/No]
[Justification in 1-3 sentences]

Example (continued from above):
Yes
The user is asking about risk factors for a medical condition, which is relevant for healthcare professionals diagnosing and treating patients. 

Do not include any other text in your response.
\end{lstlisting}
\subsection{Chat overarching user request}
This prompt instructs the model to summarize a WildChat chat into an overarching user request.
\begin{lstlisting}[style=promptstyle]
Here is a conversation between a user and an AI assistant.

<conversation>
{conversation}
</conversation>

Extract the user's main request and summarize it as an imperative command using simple, clear language. Follow these rules:

1. Start with an action verb (e.g., "Find," "Explain," "Create," "Help")
2. Use plain language instead of technical terms
3. Keep it to 3 sentences maximum
4. Focus on what the user actually wants, not the assistant's response

Example:
<conversation>
user: Old age PT hx of DM, HTN, dyslipidemia His ECG I.II, aVF (MI) what is the highest risk factor for this condition?

A: The highest risk factor for this condition (myocardial infarction, MI) in this patient would be old age combined with a history of diabetes mellitus (DM), hypertension (HTN), and dyslipidemia.
</conversation>

Response: Find the biggest risk factor for heart attacks in elderly patients with diabetes and high blood pressure.

Respond only with the imperative summary. Do not include any other text.
\end{lstlisting}
\subsection{Tool-Use Scenario Generation}
\label{app:prompt:genq}

This prompt instructs the model to generate a realistic, self-contained benchmark scenario grounded in O*NET occupational tasks. The model selects appropriate MCP server tools, designs a cross-tool workflow, and produces a user request together with the expected tool calls and outputs.

\begin{lstlisting}[style=promptstyle]
## Task
Generate a *tool-use scenario* grounded in workplace tasks performed by a given occupation.

## Objective

Brainstorm a workplace scenario in which {OCCUPATION} needs to perform *all of the following workplace tasks*, and analyze the provided MCP servers and their available tools to create
- a realistic user request that requires the use of *at least {NUM_TOOLS} target tools* to fulfill completely
- a list of target tools calls that must be made to fulfill the request and their outputs

## Workplace Tasks

**Occupation:** {OCCUPATION}
**Occupation Description:** {OCCUPATION_DESCRIPTION}
**Tasks:**
{TASKS}

Here are some search results related to the occupation and tasks. Use them to add constraints, context, and details to the request

{TASK_REFERENCES}

## MCP Servers

{SERVER_DESCRIPTIONS}

## Guidelines

### Scenario Brainstorming

- Think of realistic, specific scenarios where someone would need to use at least {NUM_TOOLS} target tools provided by the MCP servers to accomplish a meaningful task
- Consider diverse real-world contexts such as:
  - Content creators managing their online presence across different platforms
  - Researchers gathering and analyzing information from multiple sources
  - Developers building and deploying applications using different services
  - Business professionals managing projects and communications across platforms
  - Students working on complex assignments requiring multiple tools
  - Entrepreneurs launching new ventures using various services
- The scenario should be detailed and authentic, representing genuine use cases that span multiple services

### Request Realism

- Create requests that represent real-world scenarios where users would genuinely need the tools provided by the MCP servers
- The request should sound natural and authentic, as if asked by someone with a specific goal
- Include relevant context, constraints, and details that make the request engaging
- Consider workflows that require multiple complementary tools working together across different services
- Think about how different servers support each other in real-world use cases
- Use the search results to add constraints, context, and details to the request

### Request Self-Containment (Critical)

- **All information the agent needs to make tool calls must be present in the request itself.** This includes specific values such as: project names, IDs, file names, URLs, dates, usernames, account names, search queries, locations, or any other concrete parameters required by the target tools.
- The request must be **answerable only by calling the target tools** -- it should not be fulfillable from general knowledge or common sense alone. The answer depends on live data or system state that only the tools can retrieve.
- A well-formed request lets an agent immediately begin making tool calls without asking the user for clarification. If an agent receiving this request would reasonably say "I don't have enough information to proceed," the request is incomplete.
- **Do not ask vague questions** like "can you check my account status?" -- instead specify "can you check the status of account `acme-corp` and retrieve the last 5 invoices?" so the agent has concrete parameters to pass to the tools.

### Server and Target Tools Selection

- Select *at least {NUM_TOOLS} target tools* that work together
- The request should require a sequence or combination of tool calls to solve completely
- Choose target tools based on how they complement each other across different services/domains
- Consider each tool's description and purpose when crafting the cross-server workflow
- Ensure target tool calls create a logical, interconnected workflow

### Multi-Tool Integration

- Think about how different tools' capabilities can be combined
- Consider how data flows between tools and which **dependency patterns** connect them:
  - **Parameter dependency**: One tool's output provides input for the next (e.g., a lookup result feeds into a calculation)
  - **Conditional routing**: A tool's result determines which tool to call next (e.g., an inspection finding a violation triggers a reporting tool rather than a routine filing tool)
  - **Cross-validation**: Two tools verify or contradict each other's findings on the same request
  - **Aggregation**: Parallel tool calls whose results must be combined into a single response
- Create realistic scenarios where multiple tools need to work together
- Focus on complementary functionalities across different domains

### Request Complexity

- Create requests that are complex enough to warrant using at least {NUM_TOOLS} target tools across multiple servers
- The request should have multiple components or require several steps that span different services
- Include relevant context or constraints that make the multi-tool usage necessary
- Do not contain the exact target tool names or server names in the request
- Ensure the request cannot be reasonably fulfilled with tools from just a single server
- Create scenarios that naturally require different types of services working together

### Output Format

Your response should include:
1. **Tool Analysis**: Briefly analyze the tools and the workplace tasks they can help accomplish.
2. **Cross-Tool Workflow**: Describe the workflow showing how tools will be used together, including the dependencies among tools and any decision points where intermediate results affect the next step.
3. **Target Tools**: The specific tools, their server names, their input arguments, AND the output from executing the tools. The input arguments MUST follow the tool's input schema exactly (including parameter names, required fields, and value types).
4. **Request**: A clear, realistic user request that requires tool usage to accomplish the workplace tasks.

## Output
Ensure your request is grounded in all {NUM_TASKS} workplace tasks and uses at least {NUM_TOOLS} tools to solve completely. Provide your response in the following JSON format:

Machine-readable schema source of truth: `prompts/genq_from_onet_tasks_output_schema.json`.

```json
{
  "tool_analysis": "Briefly analyze the tools and how they help accomplish the workplace tasks.",
  "cross_tool_workflow": "Describe the workflow: for each tool-to-tool link, state the dependency type (parameter dependency, conditional routing, cross-validation, or aggregation) and note any decision points where intermediate results change the next step.",
  "target_tools": [
    {
      "server": "Server1",
      "tool": "get_weather",
      "arguments": {"location": "Paris, France"},
      "output": "Temperature: 18 degrees C, Conditions: Partly cloudy, Humidity: 65%"
    },
    {
      "server": "Server2",
      "tool": "send_email",
      "arguments": {"to": "bob@email.com", "body": "Hi bob"},
      "output": "Email sent successfully to bob@email.com"
    }
  ],
  "request": "A clear, realistic user request grounded in the workplace tasks that requires tool usage to fulfill. Must include all specific parameter values (names, IDs, dates, URLs, etc.) the agent needs to call the tools -- the agent should be able to start making tool calls immediately without asking for clarification."
}
```
\end{lstlisting}

\subsection{Synthetic Tool Response Simulation}
\label{app:prompt:virtual}

A ``Scenario Master'' model simulates MCP tool responses without live API access. The model receives a system prompt that defines validation rules and output format, followed by conversation context for scenario continuity, and a per-call user message specifying the tool to simulate. The full message sequence for the $k$-th tool call in a trajectory is:

\begin{enumerate}
  \item \textbf{System prompt} (Listing~\ref{lst:virtual-system}): validation rules and realism constraints.
  \item \textbf{Conversation history}: prior agent--user dialogue turns (normalized to user/assistant roles), giving the Scenario Master awareness of the overall task.
  \item \textbf{Prior tool simulation turns}: the $k{-}1$ previous tool simulation request/response pairs (as user/assistant messages), maintaining cross-tool continuity (e.g., an ID returned by tool~1 should appear in tool~2's response).
  \item \textbf{Current tool call} (Listing~\ref{lst:virtual-user}): the tool metadata and arguments for the call to simulate.
\end{enumerate}

\begin{lstlisting}[style=promptstyle,caption={Virtual tool simulation --- system prompt},label={lst:virtual-system}]
You are the Scenario Master for tool-calling scenarios, similar to a DnD dungeon master.
Your job is to co-create and play out realistic tool-calling scenarios with the user over multiple turns.
When the user provides tool simulation requests, treat them as scenario events and simulate the tool responses as the scenario unfolds.

You must always return exactly one JSON object in this schema:
{
  "error": "",
  "response": ""
}

You will receive tool metadata and tool input for each event.
Follow this process exactly:

STEP 1: STRICT VALIDATION
Special case: if the tool documentation contains no inputSchema, or inputSchema has no properties
(properties missing or empty {}), skip validation and go to Step 2.

Otherwise validate tool_input against input_schema:
1) Missing required arguments: all required keys must be present.
2) Hallucinated arguments: do not allow unknown keys not listed in properties.
3) Type mismatches: values must match declared types and enum constraints.

STEP 2: RESPONSE GENERATION
If validation fails:
- Halt simulation for that call.
- Put a concise, specific reason in "error".
- Set "response" to an empty string.

If validation succeeds:
- Keep "error" as an empty string.
- Populate "response" with meaningful, practical content that matches the tool's intended functionality.
- Maintain JSON integrity.
- If information is incomplete, still generate a useful, realistic response and fabricate plausible values when needed.

Realism requirements:
- Avoid obvious placeholders like Jane Doe, John Smith, Acme Corp, 123 Main St, example@email.com.
- Use plausible names, IDs, addresses, and domain details.
- Stay consistent with scenario context from previous conversation turns.

Context usage:
- Incorporate scenario context from the conversation history.
- Incorporate previous simulated tool-call turns to maintain continuity.
- Use the current server and tool details as ground truth for this specific call.

Output constraints:
- Return only the JSON object; no extra prose or markdown.
- Ensure the object is parseable JSON.
\end{lstlisting}

\begin{lstlisting}[style=promptstyle,caption={Virtual tool simulation --- per-call user message template},label={lst:virtual-user}]
Simulate the next tool call in this scenario.

Server info:
- server_id: {SERVER_ID}
- server_name: {SERVER_NAME}
- server_description: {SERVER_DESCRIPTION}

Tool info:
- tool_name: {TOOL_NAME}
- tool_description: {TOOL_DESCRIPTION}
- tool_input_schema_json:
{TOOL_INPUT_SCHEMA_JSON}

Tool call arguments (JSON):
{TOOL_ARGS_JSON}
\end{lstlisting}

\subsection{Withheld-Information Scenario Generation}
\label{app:prompt:genq-withheld}

A variant of the scenario generation prompt (Appendix~\ref{app:prompt:genq}) used for multi-turn benchmarks. The key difference is the \textbf{Information Withholding} section, which instructs the model to deliberately omit 1--3 key parameters from the user request. The agent must ask clarifying questions to obtain these values before making tool calls. Only the sections that differ from the base prompt are shown below.

\begin{lstlisting}[style=promptstyle]
## Task
Generate a *tool-use scenario with withheld information* grounded in workplace tasks performed by a given occupation.

## Objective

Brainstorm a workplace scenario in which {OCCUPATION} needs to perform *all of the following workplace tasks*, and analyze the provided MCP servers and their available tools to create
- a realistic user request that **deliberately omits 1-3 key parameters** an agent would need to ask about to complete the task
- a list of target tool calls that must be made to fulfill the request and their outputs
- the withheld parameters and the clarifying questions the agent should ask

[... Scenario Brainstorming, Request Realism, Server and Target Tools Selection, Multi-Tool Integration, and Request Complexity guidelines are identical to Appendix A.1 ...]

### Information Withholding (Critical for this variant)

**Override the self-containment requirement**: Deliberately omit 1-3 key parameters from the request that the agent would naturally need to ask about before calling tools.

- Choose parameters that are **naturally unknown to the AI** -- things a human user would know but might not think to specify upfront (e.g., account IDs, project names, date ranges, specific resource identifiers, usernames, file paths, API endpoints)
- Use **vague references** instead of specific values:
  - Instead of "account ID `acme-corp-2847`" -> use "my account"
  - Instead of "January 2024" -> use "last month" or "recently"
  - Instead of "project `frontend-redesign`" -> use "the project I've been working on"
  - Instead of "repository `my-org/backend`" -> use "my backend repo"
- The withheld parameters must be things the agent **cannot guess from context** -- they require explicit confirmation from the user
- The request must still be **realistic**: a real user might genuinely forget to specify these values upfront
- The agent should be able to understand the intent but must ask before calling tools with the missing parameters
- Withhold **1 to 3** parameters -- not so many that the request is incoherent

### Withheld Info and Follow-up Questions

- `withheld_info`: List the 1-3 parameters that were deliberately omitted from the request. For each, provide the parameter name, a brief description of why it's needed, and its actual value.
- `target_followup_questions`: Write the clarifying questions a well-behaved agent *should* ask the user to obtain the withheld parameters. Keep them natural and concise (one question can elicit multiple values if they're related).

### Output Format

[... identical to Appendix A.1 but with additional fields ...]

```json
{
  "tool_analysis": "...",
  "cross_tool_workflow": "...",
  "withheld_info": [
    {
      "parameter": "account_id",
      "description": "The user's account identifier needed to look up invoices",
      "value": "acme-corp-2847"
    }
  ],
  "target_followup_questions": [
    "What is your account ID or company name?"
  ],
  "target_tools": [ ... ],
  "request": "Can you check my account status and pull up my invoices from last month?"
}
```
\end{lstlisting}

\subsection{Simulated User}
\label{app:prompt:user}

This prompt drives the simulated user in multi-turn benchmark trajectories. The model acts as a realistic human user who issues the initial request and then interacts with the agent across multiple turns: providing withheld information when asked, redirecting the agent if it goes off track, and terminating the conversation when the workflow is complete.

\begin{lstlisting}[style=promptstyle]
### ROLE & OBJECTIVE
You are an **Expert User Simulator**.
You are NOT an AI Assistant. You are simulating a human user with a specific, complex goal who is testing a "Student AI's" ability to use tools correctly.

Your goal is to provide realistic user responses and guide the Student AI through a multi-turn conversation until it has correctly executed the intended tool workflow and delivered a presentable answer.

### THE SCENARIO
You are a user who knows exactly what result they want, but you need the Student AI to perform the work (calling tools) to get it.

### THE DATA (Script & Ground Truth)
The following is the "Ground Truth" data you need to execute the simulation.

<test_query>
{QUESTION}
</test_query>

<tool_analysis>
{TOOL_ANALYSIS}
</tool_analysis>

<workflow_analysis>
{WORKFLOW_ANALYSIS}
</workflow_analysis>

{WITHHELD_INFO}

<available_tools>
{TOOL_DESCRIPTIONS}
</available_tools>

### IMPORTANT -- GROUND YOUR EXPECTATIONS IN THE ACTUAL TOOLS
The `<available_tools>` section above lists the **exact tools** the Student has access to, with their real names and descriptions. Use this to calibrate your expectations:
- The tools may have generic or domain-mismatched descriptions (e.g., a "due diligence" tool repurposed for HR checks). **This is expected.** The Student is being asked to use these specific tools regardless of their original labeling.
- If the Student hesitates because a tool's description doesn't seem to match the scenario, **reassure them** that these are the correct tools for the task.
- **DO NOT** argue with the Student about what a tool does or demand they use a tool in a way that contradicts its documented interface.

### IMPORTANT -- YOU CANNOT SEE TOOL CALLS
You do **not** have visibility into the Student's tool calls or their raw outputs. When the Student calls a tool, you will only see the Student's text summary of the results -- **not** the underlying function call or the tool's JSON response.

This means:
- If the Student says "I searched for X and found Y," **trust that the tool call happened**.
- You have **no way** to verify whether a tool was invoked -- you can only judge the Student's final text response.
- **NEVER** say things like "you need to actually invoke the tools," "show me the real output," or "stop generating fake responses."
- Judge the Student's work based on: (a) whether their described workflow matches the expected tools and sequence, and (b) whether their final answer is coherent and addresses your request.

### INTERACTION LOGIC

#### Step 1: The Request (First Turn)
Output the content inside `<test_query>` exactly as written. Do not add extra text.

#### Step 2: The Evaluation Loop (Subsequent Turns)
Every time the Student AI responds, evaluate whether it has completed the intended workflow and provided a presentable answer.

**IF the Student has completed the workflow and given a presentable response:**
   - Reply with exactly: `"<END_CONVERSATION>"`

**IF the Student asks for clarifying information listed in `<withheld_information>` (when present):**
   - Provide the withheld value naturally, as a real user would respond.
   - Do not acknowledge that it was deliberately withheld. Stay in character.

**IF the Student is going down the wrong path or seems stuck:**
   - Gently redirect them. For example:
     - "That doesn't seem quite right -- have you considered a different approach?"
     - "I think you might need to do X before Y."
   - Base your redirection on the expected workflow, but do NOT copy-paste from ground truth sections.

**CRITICAL -- Tool output content is NOT the Student's fault:**
   - The Student has no control over what a tool returns. If a tool returns unexpected data, that is NOT a reason to challenge the Student.
   - **DO NOT** tell the Student the tool output is wrong or ask them to re-run the same tool expecting different results.

### WORKFLOW COMPLETION CHECK

End the conversation when the Student has clearly executed the intended tool workflow and delivered a coherent, presentable response. Specifically:

1. **Workflow Adherence**: The Student called the tools described in `<tool_analysis>`, following the sequence and dependency logic described in `<workflow_analysis>`.
2. **Tool Correctness**: The tools called must match the target tools -- not hallucinated alternatives.
3. **Presentable Response**: The Student provided a clear, coherent answer that addresses the original query.
4. **Ignore Tool Output Quality**: Do NOT factor in whether the tool outputs look "correct" or match your expectations.

### NEGATIVE CONSTRAINTS (CRITICAL)
- **DO NOT** reveal that you are an AI, a Simulator, or a Proctor. Stay in character as the User.
- **DO NOT** copy-paste the content of `<tool_analysis>` or `<workflow_analysis>` directly to the student.
- **DO NOT** make tool calls or ping MCP servers for the agent.
- **DO NOT** proactively reveal withheld information -- only provide withheld values when the agent explicitly asks for them.
- **DO NOT** take on the role of the Student AI. Never call tools, generate tool outputs, or do the Student's work.
\end{lstlisting}

\subsection{Tool Call Accuracy Evaluation}
\label{app:prompt:evaluator}

This prompt instructs the model to compare an agent's actual tool call arguments against expected arguments from an answer key. It supports both single-call and multi-call evaluation, allowing for functional equivalence (e.g., paraphrasing, case-insensitive matching) rather than requiring exact string matches.

\begin{lstlisting}[style=promptstyle]
## ROLE AND OBJECTIVE ##
You are an evaluator of AI assistants. Your job is to evaluate the assistant's tool calls against expected tool calls from an answer key.

You will be provided with the expected tool call(s) and the assistant's actual tool call(s). You must return a JSON object with your evaluation.

## WORKFLOW ##

### Single-Call Evaluation
When a single expected and actual call are provided:
1. Compare the assistant's tool call arguments to the expected arguments.
2. A correct tool call should either match the answer exactly OR be functionally the same (for example, a re-wording of a 'key phrase' argument, or case-insensitive matching).
3. Score 1 if correct, 0 if incorrect.

### Multi-Call Evaluation
When multiple expected and actual calls are provided for the same tool:
1. Determine whether each expected call has a functionally equivalent match among the actual calls.
2. Order does NOT matter -- the agent may have called the tool in a different sequence.
3. A match follows the same criteria as single-call: exact match or functionally equivalent arguments.
4. Score 1 if ALL expected calls have a match among the actual calls, 0 if any expected call lacks a match.

## OUTPUT FORMAT ##
You must return a JSON object with the following fields:
{
  "reasoning": "A brief explanation of why the call(s) are correct or incorrect.",
  "score": 1 (if correct) or 0 (if incorrect)
}

## EXAMPLES ##

### Single-Call Example

EXPECTED Tool: use_computer
EXPECTED Arguments: {"prompt": "Enter a new patient record with the following details: Name: Jane Smith, DOB: 1975-06-22, Sex: Female, Diagnosis: Type 2 Diabetes, Medications: Metformin 500 mg twice daily, Allergies: Penicillin."}

ACTUAL Tool: use_computer
ACTUAL Arguments: {"prompt": "Add a new patient record with the following details:\nName: Jane Smith\nDOB: 1975-06-22\nSex: Female\nDiagnosis: Type 2 diabetes\nMedications: Metformin 500 mg twice daily\nAllergies: Penicillin.\nPlease confirm the entry was successful."}

Evaluation:
{
    "score": 1,
    "reasoning": "The assistant invoked `use_computer` with a prompt that includes all required patient details (name, DOB, sex, diagnosis, medications, allergies). Although the wording differs slightly from the answer key, the content is functionally equivalent."
}

### Multi-Call Example

EXPECTED Calls for tool `search_employee` (order does NOT matter):

Expected Call 1: {"search_query": "engineering department managers", "search_type": "auto"}
Expected Call 2: {"search_query": "HR benefits coordinator", "search_type": "auto"}

ACTUAL Calls for tool `search_employee` (order does NOT matter):

Actual Call 1: {"search_query": "HR benefits coordinator contact", "search_type": "auto"}
Actual Call 2: {"search_query": "engineering dept managers", "search_type": "auto"}

Evaluation:
{
    "score": 1,
    "reasoning": "Both expected calls have functional matches: Expected Call 1 ('engineering department managers') matches Actual Call 2 ('engineering dept managers') -- same intent. Expected Call 2 ('HR benefits coordinator') matches Actual Call 1 ('HR benefits coordinator contact') -- same target with minor extra wording. Order differs but all expected calls are covered."
}
\end{lstlisting}

\subsection{Workflow Completion Evaluation}
\label{app:prompt:workflow}

This prompt instructs the model to evaluate whether an agent followed the expected cross-tool workflow: the correct tool ordering, proper inter-tool data passing, and coherent final synthesis. Responses are scored on a five-point scale from \textit{No Workflow} (1) to \textit{Complete Workflow} (5).

\begin{lstlisting}[style=promptstyle]
## ROLE AND OBJECTIVE ##
You are an evaluator of AI assistants that use tools to accomplish multi-step workplace tasks. Your job is to evaluate whether the assistant followed the **expected cross-tool workflow** -- the intended sequence of tool calls, inter-tool data flow, and final synthesis.

You will be provided with:
1. The original question given to the assistant
2. The expected cross-tool workflow (reference describing the intended tool sequence and how outputs should connect)
3. Supporting tool analysis (describing how each tool contributes to the tasks)
4. A condensed version of the assistant's full trajectory (tool calls, tool outputs, and assistant messages)
5. The assistant's final response

## WORKFLOW ##

1. Read the **expected cross-tool workflow** carefully to understand the intended tool sequence and data flow.
2. Walk through the **agent trajectory** and map the actual tool calls against the expected sequence.
3. Assess three aspects:
   - **Tool ordering**: Did the agent call tools in the correct sequence?
   - **Data flow**: Did the agent use outputs from earlier tools as inputs to later tools (as described in the workflow)?
   - **Synthesis**: Did the agent produce a coherent final response that integrates results from the workflow steps?

## SCORING SCALE ##

| Score | Label          | Meaning                                                                                         |
|-------|----------------|-------------------------------------------------------------------------------------------------|
| 1     | No Workflow    | No relevant tool calls were made. The agent did not attempt the expected workflow.              |
| 2     | Wrong Workflow | Tools were called but in the wrong order, wrong tools were used, or outputs were not passed between steps as expected. |
| 3     | Partial Workflow | Some workflow steps were executed correctly, but key stages are missing or disconnected (e.g., skipped a critical intermediate step). |
| 4     | Mostly Complete | The correct tool sequence was followed with proper data flow, but there are minor gaps -- e.g., a non-critical step was skipped or the final synthesis is incomplete. |
| 5     | Complete Workflow | The full workflow was executed: correct tool ordering, proper inter-tool data passing, and a coherent final response that integrates all workflow results. |

## GUIDELINES ##

- **Focus on workflow structure, not content quality**: You are evaluating whether the agent followed the right process, not whether its final answer is factually perfect.
- **Alternative but equivalent workflows are acceptable**: If the agent achieved the same data flow using different but functionally equivalent tool calls, give credit.
- **Partial credit for partial workflows**: If the agent completed 3 out of 4 expected steps correctly, that is "Mostly Complete" not "Wrong Workflow".
- **Tool call failures count against the workflow**: If a tool call was made but returned an error, that step is incomplete.
- **The reference workflow is a guide, not a rigid script**: Minor deviations in ordering that don't affect the logical data flow should not be penalized heavily.
- **Extra tool calls are fine**: Additional tool calls beyond what was expected do not reduce the score, as long as the core workflow was followed.

## OUTPUT FORMAT ##
You must return a JSON object with the following fields:
{
  "expected_steps": ["Step 1: ...", "Step 2: ...", "..."],
  "actual_steps": ["What the agent actually did for step 1", "..."],
  "reasoning": "Analysis of how well the actual workflow matched the expected workflow",
  "rating": "no workflow | wrong workflow | partial workflow | mostly complete | complete workflow",
  "score": 1-5
}
\end{lstlisting}

\subsection{Response Grounding Evaluation}
\label{app:prompt:grounding}

This prompt instructs the model to assess whether an agent's factual claims are supported by the tool outputs it actually received. Each claim is classified as grounded or ungrounded, and an overall score from \textit{Ungrounded} (1) to \textit{Fully Grounded} (5) is assigned.

\begin{lstlisting}[style=promptstyle]
## ROLE AND OBJECTIVE ##
You are an evaluator of AI assistants that use tools to answer questions. Your job is to evaluate whether the assistant's claims are **grounded** -- that is, supported by the tool outputs the assistant actually received during the conversation.

You will be provided with:
1. The original question given to the assistant
2. All tool call / tool output pairs from the trajectory (the evidence base)
3. All assistant messages from the trajectory (the claims to evaluate)

## WORKFLOW ##

1. Review all **tool outputs** -- these are the facts available to the assistant.
2. Review all **assistant messages** (intermediate and final) -- these are the claims to check.
3. For each factual claim the assistant makes, determine whether it is supported by a tool output the assistant received.
4. Compile lists of grounded and ungrounded claims.
5. Assign an overall grounding score.

## SCORING SCALE ##

| Score | Label              | Meaning                                                                        |
|-------|--------------------|--------------------------------------------------------------------------------|
| 1     | Ungrounded         | Mostly fabricated; no meaningful connection to tool outputs.                   |
| 2     | Poorly Grounded    | Some claims connect to tool outputs, but the majority are unsupported or distorted. |
| 3     | Partially Grounded | Key claims are grounded, but there are notable unsupported assertions.         |
| 4     | Mostly Grounded    | The vast majority of claims are traceable to tool outputs; minor inferences are acceptable. |
| 5     | Fully Grounded     | All factual claims are directly supported by tool outputs.                     |

## GUIDELINES ##

- **Focus on factual claims**: numbers, names, dates, data points, specific results. Ignore stylistic elements, greetings, or conversational filler.
- **Reasonable summarization is acceptable**: The assistant paraphrasing or summarizing tool output content is grounded, as long as the meaning is preserved.
- **Error acknowledgment is grounded**: If the assistant says "I couldn't find X" and the tool returned an error or empty result, that IS grounded.
- **Generic knowledge is neutral**: Statements like "This tool helps with X" or general domain knowledge are neither grounded nor ungrounded -- do not penalize or reward them.
- **No tool calls + substantive claims = likely ungrounded**: If the assistant provided detailed factual information without making any tool calls, it is likely fabricating.
- **Check against the assistant's OWN tool outputs**: The grounding check is about whether claims match what the assistant received, NOT whether the claims are objectively true.

## OUTPUT FORMAT ##
You must return a JSON object with the following fields:
{
  "grounded_claims": ["claim 1 backed by tool output X", "claim 2 supported by ..."],
  "ungrounded_claims": ["claim 3 not supported by any tool output", "..."],
  "reasoning": "Overall assessment of the assistant's grounding",
  "rating": "ungrounded | poorly grounded | partially grounded | mostly grounded | fully grounded",
  "score": 1-5
}
\end{lstlisting}
\subsection{Autonomy Evaluation}
\label{app:prompt:autonomy}

This prompt instructs the model to score how independently the agent completed the multi-turn workflow. Essential clarification questions (e.g., requesting unknown parameter values) are not penalised; only unnecessary user steering reduces the score. Responses are scored on a five-point scale from \textit{Did Not Complete} (1) to \textit{Autonomous Completion} (5).

\begin{lstlisting}[style=promptstyle]
## ROLE AND OBJECTIVE ##
You are an evaluator of AI assistants that solve multi-turn tool-use tasks.
Your job is to score the assistant's **autonomy**: how independently it completed the workflow while only asking for essential, a priori unknown user information.

You will be provided with:
1. The original question
2. The expected cross-tool workflow
3. Optional tool-analysis context
4. Optional withheld-information reference
5. Optional target follow-up questions reference
6. A condensed trajectory (assistant + user + tool interactions)
7. A telemetry summary (turn counts and intervention indicators)

## WHAT AUTONOMY MEANS ##
Autonomy is high when the assistant drives execution itself, asks only necessary clarifying questions, and completes the workflow without repeated user steering.

Do NOT penalize the assistant for asking for truly missing, essential information that could not have been known in advance.

## SCORING SCALE (1-5) ##

| Score | Label | Meaning |
|-------|-------|---------|
| 1 | did not complete | Workflow was not completed (including turn-max / turn-expired), or required heavy user steering and still failed to finish. |
| 2 | heavy assistance | Completed little of the workflow without substantial user redirection; user had to repeatedly steer core execution. |
| 3 | moderate assistance | Partial-to-mostly completion, but multiple meaningful user interventions were needed to correct or unblock progress. |
| 4 | mostly autonomous | Workflow completed with limited user steering; clarifications were mostly essential and execution was largely self-directed. |
| 5 | autonomous completion | Workflow completed end-to-end autonomously, except for essential unknown inputs the user needed to provide. |

## GUIDELINES ##
- Distinguish between:
  - **Essential clarification**: required unknown values (acceptable)
  - **Steering intervention**: user correcting order, re-running, redirecting wrong path, or compensating for assistant confusion
- Treat **target follow-up questions as a non-exhaustive reference**, not a strict checklist.
- Do not penalize extra clarification questions when they are necessary to execute the workflow, even if they are not listed in target follow-up questions.
- If the trajectory clearly did not complete, score 1.
- Use trajectory evidence and telemetry; telemetry supports but does not override trajectory evidence.
- Focus on workflow execution autonomy, not prose style.

## OUTPUT FORMAT ##
Return only JSON:
{
  "score": <integer 1-5>,
  "rating": "did not complete | heavy assistance | moderate assistance | mostly autonomous | autonomous completion",
  "reasoning": "2-4 sentences citing concrete trajectory behavior and user intervention patterns."
}
\end{lstlisting}

\subsection{Follow-up Quality Evaluation}
\label{app:prompt:followup}

This prompt instructs the model to evaluate whether the agent correctly identified deliberately withheld parameters and asked targeted clarifying questions before making tool calls. It is only applied to scenarios with withheld information. Responses are scored on a five-point scale from \textit{Skipped Clarification} (1) to \textit{Complete Clarification} (5).

\begin{lstlisting}[style=promptstyle]
## Role
You are an expert evaluator assessing whether an AI agent correctly identified and asked for missing information before attempting to use tools.

## Task
The user's request deliberately omitted key parameters that the agent needed to obtain through clarifying questions. Your job is to evaluate whether the agent asked the right follow-up questions and successfully used the provided information in its tool calls.

## Inputs

**Original Request (with deliberate omissions):**
The request uses vague references (e.g., "my account", "the project") where specific values are required.

**Withheld Information:**
The parameters that were deliberately omitted, along with their correct values.

**Target Follow-up Questions:**
The clarifying questions a well-behaved agent should have asked.

**Agent Trajectory:**
The full conversation, including the agent's questions, the user's answers, and any tool calls made.

---

## Scoring Rubric (1-5)

**Score 5 -- Complete Clarification**
The agent identified all withheld parameters, asked clear and targeted questions, received the values from the user, and correctly used them in subsequent tool calls. The clarification felt natural and efficient (minimal back-and-forth).

**Score 4 -- Complete with Minor Issues**
The agent asked for all required withheld parameters but with some inefficiency: awkward phrasing, extra clarification turns for information that could have been combined, or slight redundancy. All withheld values were ultimately obtained and used correctly.

**Score 3 -- Partial Clarification**
The agent asked for some but not all withheld parameters, OR asked the right questions but missed one key parameter, requiring re-prompting. The agent may have made some tool calls with placeholder/guessed values, or the conversation needed extra turns due to incomplete questioning.

**Score 2 -- Vague or Irrelevant Questions**
The agent asked questions but they were too vague to elicit the specific withheld values, OR the questions were tangential and did not target the missing parameters. The agent may have eventually obtained some information through repeated prompting but failed to identify the core gap.

**Score 1 -- Skipped Clarification**
The agent proceeded with tool calls without asking for the withheld parameters, using guessed values, placeholder values, or simply failing. The agent showed no attempt to identify that critical information was missing.

---

## Output Format

```json
{
  "score": <integer 1-5>,
  "rating": "<one of: complete clarification | complete with minor issues | partial clarification | vague or irrelevant questions | skipped clarification>",
  "reasoning": "<2-4 sentences explaining the score. Cite specific agent behavior: what it asked, what it missed, and whether withheld values were correctly used in tool calls.>"
}
```

Respond with only the JSON block above -- no additional text.
\end{lstlisting}

\subsection{WildChat Task Filtering}
\label{app:prompt:filter}

This prompt is used during data preparation (Section 2) to filter pre-assigned O*NET tasks for each WildChat chat. Each chat enters this stage with a brief summary and three candidate occupational tasks selected by an upstream embedding-based mapping. 
\begin{lstlisting}[style=promptstyle,caption={WildChat task filtering --- system prompt},label={lst:filter-system}]
You are an expert classifier matching AI chat summaries to occupational tasks. IMPORTANT: The summaries are very brief and may omit key details. Tasks are described in formal occupational language that may sound different from casual AI chat topics, but can still be the same underlying activity. For example, 'calculate equilibrium constant Kp' matches 'calculate amounts of chemicals using mathematical formulas' -- the specific domain differs but the core skill is the same. Be INCLUSIVE: keep any task where the core skill or activity overlaps, even if the domain or wording differs. Only reject tasks that have NO conceivable connection to the summary.
\end{lstlisting}

\begin{lstlisting}[style=promptstyle,caption={WildChat task filtering --- per-chat user message template},label={lst:filter-user}]
A user had this conversation with an AI assistant (this is only a brief summary -- the actual conversation likely covered more ground):
"{summary}"

These occupational tasks were identified as potentially related. A task matches if the underlying skill or activity overlaps, even if the specific domain or terminology differs:
1. {task1}
2. {task2}
3. {task3}

Which tasks have any overlap with the conversation's activities? Return ONLY a comma-separated list of task numbers (e.g. "1,2,3") or "None" if absolutely none are related.
\end{lstlisting}

\subsection{MCP Server Quality Assessment}
\label{app:prompt:serverqual}

This prompt is used during MCP server filtering (Appendix~A) to assess the quality of a candidate server's tools before it is admitted to the benchmark pool. The agent is given the server name and a formatted list of its tools (each with its name, description, and input schema) and is instructed to call every tool with the most realistic arguments it can construct, retry on input-related errors, and emit a structured pass/fail verdict per tool. 
\begin{lstlisting}[style=promptstyle]
You have access to the '{SERVER_NAME}' MCP server. Your job is to test each tool listed below and then report the quality of each tool's output.

Tools to test:
{TOOL_LIST}

Instructions:
1. Call EACH tool at least once using the most realistic inputs you can construct based on the tool's description and input schema
2. If a tool returns an error caused by your inputs (e.g. invalid argument value, missing required field), retry with corrected inputs (up to 2 more attempts per tool)
3. If a tool fails due to a server or connection error (e.g. 500, timeout, unreachable), note the failure and move on to the next tool
4. After testing ALL tools, output ONLY a JSON object as your final message -- no markdown fences, no explanation, just the raw JSON:

{
  "tool_results": [
    {
      "tool_name": "<exact tool name>",
      "quality": "<pass or fail>",
      "reasoning": "<brief explanation of what the tool returned and why it passes or fails>"
    }
  ]
}

"quality" must be "pass" if the tool returned real, non-error data consistent with its described purpose.
"quality" must be "fail" if the tool only returned errors on all attempts, was unreachable, or returned empty/meaningless output.
\end{lstlisting}



\end{document}